\documentclass[onecolumn]{IEEEtran}
\usepackage{amsfonts,amsmath,amssymb,amsthm,mathrsfs}
\usepackage{graphicx,multirow,bm}
\usepackage{color}
\usepackage[noadjust]{cite}
\usepackage{refcheck}
\makeatletter
\newtheoremstyle{mythm}{3pt}{3pt}{}{16pt}{\bfseries}{:}{.5em}{}
\theoremstyle{mythm}
\newtheorem{theorem}{Theorem}
\newtheorem{definition}{Definition}
\newtheorem{remark}{Remark}

\newtheorem{proposition}{Proposition}
\newtheorem{corollary}{Corollary}
\newtheorem{lemma}{Lemma}

\newtheorem{construction}{Construction}

\newcommand{\balpha}{\boldsymbol{\alpha}}
\newcommand{\btheta}{\boldsymbol{\theta}}

\newcommand{\cA}{\mathcal{A}}

\newcommand{\cC}{\mathcal{C}}

\newcommand{\cE}{\mathcal{E}}

\newcommand{\cI}{\mathcal{I}}

\newcommand{\cR}{\mathcal{R}}

\newcommand{\bA}{{\bm A}}

\newcommand{\bH}{{\bm H}}

\newcommand{\bS}{{\bm S}}

\newcommand{\bY}{{\bm Y}}

\newcommand{\bma}{{\bm a}}
\newcommand{\bmb}{{\bm b}}

\newcommand{\bms}{{\bm s}}

\newcommand{\N}{\mathbb{N}}

\newcommand{\F}{\mathbb{F}}

\DeclareMathOperator{\spn}{span}
\DeclareMathOperator{\rank}{rank}

\DeclareMathOperator{\roots}{Roots}
\DeclareMathOperator{\tr}{Tr}
\renewcommand{\le}{\leqslant}
\renewcommand{\leq}{\leqslant}

\renewcommand{\geq}{\geqslant}

\newcommand{\abs}[1]{\left|#1\right|}

\newcommand{\parenv}[1]{\left( #1 \right)}

\newcommand{\bracenv}[1]{\left\{#1 \right\}}

\newcommand{\eqdef}{\triangleq}
\newcommand{\GRS}{\mathrm{GRS}}

\usepackage{pgf}
\usepackage{tikz}
\usetikzlibrary{matrix,chains,positioning,arrows,calc,decorations,plotmarks,patterns,trees,fit,backgrounds,shapes.multipart,topaths}
\usepackage{pgfplots}
\usetikzlibrary{external}                       
\usepackage{scalefnt}
\usepackage{tkz-graph}
\pgfplotsset{compat=1.3}
\tikzstyle{help lines}=[black!20,dashed]

\pgfplotscreateplotcyclelist{mylist}{red,blue,black,yellow,brown}

\pgfdeclarepatternformonly{my north east lines}{\pgfqpoint{-1pt}{-1pt}}{\pgfqpoint{8pt}{8pt}}{\pgfqpoint{6pt}{6pt}}%
{
	\pgfsetlinewidth{0.4pt}
	\pgfpathmoveto{\pgfqpoint{0pt}{0pt}}
	\pgfpathlineto{\pgfqpoint{6.1pt}{6.1pt}}
	\pgfusepath{stroke}
}
\definecolor{light_gray}{rgb}{0.6,0.6,0.6}
\definecolor{awgray}{rgb}{0.7,0.7,0.7}
\definecolor{awgray_dark}{rgb} {0.4,0.4,0.4}

\tikzset{
	>=stealth',
	mycircle/.style={circle, draw=gray, very thick},
	mycircle_small/.style={circle,draw=awgray_dark,fill = awgray_dark, inner sep=0,minimum size=.6em},
	mycircle_small_black/.style={circle,draw=black,fill = black, inner sep=0,minimum size=.6em},
	mybox/.style={rectangle,rounded corners,draw=black, thick,text width=1em,minimum height=4em,minimum width=4em,text centered},
	mybox_small/.style={rectangle,rounded corners,draw=black, thick,text width=1em,minimum height=2em,minimum width=2em,text centered},
	mybox_vec/.style={rectangle,rounded corners,draw=black, thick,text width=1em,minimum height=0.7em, minimum width=4em,text centered},
	mybox_vec_short/.style={rectangle,rounded corners,draw=black, thick,text width=1em,minimum height=0.7em, minimum width=2em,text centered},
	pil/.style={->, thick, shorten <=2pt, shorten >=2pt,},
}

\begin{document}

\title{Repairing Schemes for Tamo-Barg Codes}
\author{Han Cai, \IEEEmembership{Member,~IEEE}, Ying Miao, Moshe Schwartz, \IEEEmembership{Fellow,~IEEE},\\ and Xiaohu Tang, \IEEEmembership{Senior Member,~IEEE}
\thanks{The material in this paper was submitted in part to the IEEE International
Symposium on Information Theory (ISIT 2024).}
\thanks{H. Cai and X. Tang are with the School of Information Science and Technology,
  Southwest Jiaotong University, Chengdu, 610031, China (e-mail: hancai@aliyun.com;xhutang@swjtu.edu.cn).}
\thanks{Y. Miao is with the Faculty of Engineering, Information and Systems,
University of Tsukuba, Tennodai 1-1-1, Tsukuba 305-8573, Japan (e-mail: miao@sk.tsukuba.ac.jp).}
\thanks{M. Schwartz is with the Department of Electrical and Computer Engineering, McMaster University, Hamilton, Ontario, L8S 4K1, Canada, and on leave of absence from the School of Electrical and Computer Engineering, Ben-Gurion University of the Negev, Israel (e-mail: schwartz.moshe@mcmaster.ca).}
}
\maketitle

\begin{abstract}
{In this paper, the repair problem for erasures beyond locality in locally repairable codes is explored under a practical system setting, where a rack-aware storage system consists of racks, each containing a few parity checks. This is referred to as a rack-aware system with locality. Two repair schemes are devised to reduce the repair bandwidth for Tamo-Barg codes under the rack-aware model by setting each repair set as a rack. Additionally, a cut-set bound for locally repairable codes under the rack-aware model with locality is introduced. Using this bound, the second repair scheme is proven to be optimal. Furthermore, the partial-repair problem is considered for locally repairable codes under the rack-aware model with locality, and both repair schemes and bounds are introduced for this scenario.}

\end{abstract}


\section{Introduction}

\IEEEPARstart{W}{ith} the expanding volume of data in large-scale
cloud storage and distributed file systems like Windows Azure Storage
and Google File System (GoogleFS), disk failures have become a norm
rather than an exception. To protect data from such failures, the
simplest solution is to replicate data packets across different
disks. However, this approach suffers from large storage
overhead. Consequently, coding techniques have been developed as an
alternative solution.

Several performance metrics have been introduced to assess the codes'
performance, taking into account various aspects of the storage system
that are of interest. In order to maximize failure tolerance and
minimize redundancy, \emph{maximum distance separable (MDS) codes}
have been considered. To minimize the number of bits communicated
during the repair procedure, codes called \emph{regenerating codes
with optimal repair bandwidth} have been
developed~\cite{dimakis2010network}.  At the same time, \emph{locally
repairable codes} have been proposed{ ~\cite{gopalan2012locality}}, to reduce the number of nodes
participating in the repair process. To ensure that data can be
frequently accessed by multiple processes in parallel, codes that
support parallel reads were
introduced~\cite{wang2014repair,rawat2016locality}. Finally, to
improve the update and access efficiency, codes with optimal access
and update properties have also been
considered~\cite{tamo2012zigzag,mazumdar2014update}.  Over the past
decade, many results have been obtained regarding codes for
distributed storage systems according to these metrics, e.g., see
\cite{dimakis2010network,guruswami2017repairing,tamo2018repair,ye2017explicit,tamo2017optimal,
li2018generic,li2016optimal,tamo2012zigzag,zeh2016bounds,liu2023generic,hu2017optimal,hou2019rack,chen2020explicit,wang2023rack,zhang2023vertical}
for codes with optimal repair bandwidth,
see~\cite{huang2013pyramid,gopalan2012locality,rawat2013optimal,tamo2014family,cadambe2015bounds,kim2018locally,
li2019construction,xing2019construction,cai2020optimal,cai2021optimal_GPMDS,chen2020improved,hao2020bounds,sasidharan2015codes}
for optimal locally repairable codes,
see~\cite{rawat2016locality,TaBaFr16bounds,cai2018optimal,cai2019optimal,jin2019construction}
for codes with good availability, and
see~\cite{tamo2012zigzag,mazumdar2014update,li2015framework,chen2020enabling}
for codes with optimal update and access properties.

In this paper, we focus on locally repairable codes and codes with
optimal repair bandwidth. For a locally repairable code, all code
symbols are partitioned into repair sets, each containing some
redundancy to allow local repair. When a small prescribed number of
erasures affect a repair set, the repair process is designed to be as
easy as repairing a short MDS code, usually accessing far less data
than the amount of the original data encoded into the entire
codeword. However, when the erasure patterns exceed the local-repair
ability, the repair problem is still open, which may result in a
repair scheme similar to that of MDS codes, requiring a bandwidth as
large as the original data. { Our motivation is to consider the repair
problem for erasures beyond locality for locally repairable
codes, i.e., to find an efficient repair scheme for this kind of erasure patterns.} 

In the literature, a known approach to combine locality and
regenerating codes is to include redundancies in each repair set,
allowing the codes in each set to form regenerating codes, e.g.,
\cite{kamath2014codes,holzbaur2021partial,li2022pmds}. By doing so, the repair
bandwidth required can be reduced when the system performs local
erasure repairs.

However, this method has a drawback: the repair property only works
for the punctured codes in the repair sets. This means that if there
are erasures beyond the local repair capability in one repair set, the
repair scheme and the locality cannot simultaneously reduce the repair
bandwidth. To address this issue, we propose a new combination
strategy to repair erasures beyond the locality. Our approach involves repair schemes for locally repairable
codes that can handle erasures beyond local recoverability. In addiction, from a
practical perspective, our idea is motivated by the observation that
repair sets may be located on the same server or physically nearby
servers, and communication within a repair set may be less expensive
than communication across repair sets, which is the practical setting for
rack-aware storage systems. Therefore, desirable repairing
schemes should be able to reduce the bandwidth required for
communication across repair sets.

Specifically, in this paper, we propose repair schemes for the
well-known Tamo-Barg codes~\cite{tamo2014family}, which are optimal
locally repairable codes with respect to the Singleton-type
bound~\cite{gopalan2012locality,prakash2012optimal}. We present two
proposed schemes. Firstly, in a rack-aware model where each repair set
is one rack, we introduce an optimal repair scheme for the case of one
failed rack, i.e., one erased repair set. Secondly, for the scenario
where there are erasures within a repair set that cannot be recovered
locally, we introduce a repair scheme that reduces the repair
bandwidth required for recovering those failures. We prove the
optimality of our schemes by modifying the well-known cut-set
bound~\cite{dimakis2010network} to incorporate locality. Our proposed
schemes generalize the rack-aware model regenerating
codes~\cite{hu2016double,hou2019rack}.

The remainder of this paper is organized as
follows. Section~\ref{sec-preliminaries} introduces some preliminaries
about locally repairable codes and regenerating codes.
Section~\ref{sec-rack-aware-system} introduces the basic system setting for
rack-aware systems with locality. Section \ref{sec_TB_CRT}
 describes some basic properties for
Tamo-Barg codes.
Section~\ref{sec-TBcode-RRC} introduces schemes for repairing a whole rack erasure
(repair set).  Section~\ref{sec-partial} presents a partial repair
scheme for Tamo-Barg codes, which is capable of repairing an erased
fraction of { racks}. Additionally, the scheme is proved to
be optimal.  Section~\ref{sec-conclusion} concludes this paper with
some remarks.

\section{Preliminaries}\label{sec-preliminaries}

We start by introducing basic { notation} and definitions. For any
$n\in\N$ we denote $[n]\eqdef\{1,2,\dots,n\}$. For a prime power $q$,
let $\F_q$ denote the finite field of size $q$, $\F_q^{*}\eqdef
\F_{q}\setminus\{0\}$, and let $\F_q[x]$ denote the set of polynomials
in the indeterminate $x$ with coefficients from $\F_q$. An $[n,k]_q$
linear code $\cC$ over $\F_q$ is a $k$-dimensional subspace of
$\F_q^n$ with a $k\times n$ generator matrix
$G=(\mathbf{g}_1,\mathbf{g}_2,\dots,{\bf g}_{n})$, where
$\mathbf{g}_i$ is a column vector of length $k$ for all
$i\in[n]$. More specifically, it is called an $[n,k,d]_q$ linear code
if its minimum Hamming distance is $d$. For a subset $S\subseteq [n]$,
we use $\spn(S)$ to denote the linear space spanned by $\{\mathbf{g}_i
~:~ i\in S\}$ over $\F_q$, and $\rank(S)$ to denote the dimension of
$\spn(S)$.

\subsection{Generalized Reed-Solomon codes}

Let $\btheta=(\theta_1,\dots,\theta_n)\in\F_q^n$ contain distinct
entries, where we assume $q\geq n$. Then the well-known
\emph{generalized Reed-Solomon (GRS) code} with parameters
$[n,k,n-k+1]_q$ can be defined as
\begin{equation*}
\GRS_k(\btheta,\balpha)\eqdef\bracenv{(\alpha_1f(\theta_1),\alpha_2f(\theta_2),\dots, \alpha_n f(\theta_n)) ~:~ f(x)\in \F_q[x]\,\text{with}\,\deg(f(x))< k},
\end{equation*}
where $\balpha=(\alpha_1,\alpha_2,\dots,\alpha_n)\in (\F^*_q)^n$. It is well known that the dual of an $[n,k,n-k+1]_q$ GRS code is an $[n,n-k,k+1]_q$ GRS code (e.g., see~\cite{MaSl77theory}).

\subsection{Regenerating codes}

An important problem in distributed storage systems is to repair an
erasure by downloading as little data as possible.  Dimakis et
al.~\cite{dimakis2010network} introduced \emph{repair bandwidth}, the
amount of data downloaded during a node repair, as a metric to measure
the procedure's efficiency.

\begin{definition}
Let $\cC$ be an $[N,K]_q$ array code with \emph{sub-packetization}
$L$, that is, $c_i \in \F_q^L$ for any codeword $(c_1,c_2,\ldots,c_N)
\in \cC$.  For an erasure pattern $\cE\subseteq [N]$ and a $D$-subset
$\cR\subseteq [N]\setminus\cE$ (whose entries are called \emph{helper
nodes}), define $B(\cC,\cE,\cR)$ as the \emph{minimum repair
bandwidth} for $c_i \in \F_q^L$ stored in node $i\in\cE$, i.e., the
smallest total number of symbols of $\F_q$ helper nodes need to send
in order to recover $c_i$ (where each helper node $j\in \cR$ may send
symbols that depend solely on $c_j\in\F_q^L$).
\end{definition}

{ \begin{definition}\cite{pless1998handbook}
Let $\cC$ be an $L\times N$ array code over $\F_q$, i.e., $\cC\subseteq \F^{L\times N}_{q}.$ Then the
dimension of the array code is defined as $K=\log_{q^L}(|\cC|)$. Furthermore, the code is said to be
\emph{maximum distance separable (MDS)} code if
$$\min \{d_c(C_1,C_2)~:~C_1\ne C_2, \text{ and } C_1,C_2\in \cC\}=N-K+1,$$
where the column Hamming distance is defined as $$d_c(C_1,C_2)=|\{ i\in[N] ~:~C_{1,i}\ne C_{2,i}, C_1=(C_{1,1},C_{1,2},\cdots, C_{1,N}), \text{ and }C_2=(C_{2,1},C_{2,2},\cdots, C_{2,N})\}|.$$
\end{definition}}

In \cite{dimakis2010network}, the
well-known cut-set bound was first derived for the minimum download
bandwidth.

\begin{theorem}[Cut-set bound, \cite{dimakis2010network,cadambe2013asymptotic}]\label{theorem_cut_B}
Let $\cC$ be an $[N,K]_q$ MDS array code with sub-packetization $L$.
Let $D$ be an integer with $K\leq D\leq N-1$. For any non-empty
$\cE\subseteq [N]$ with $|\cE|\leq N-D$ and any $D$-subset
$\cR\subseteq [N]\setminus\cE$, we have
\[
B(\cC,\cE,\cR)\geq \frac{DL}{D-K+|\cE|}.
\]
\end{theorem}


\begin{definition}
For $K<D\leq N-{\tau}$, an $[N,K]_q$ MDS array code
is said to be an $[N, K]_q$ \emph{minimum storage regenerating} (MSR)
code with repair degree $D$, if for each $\cI=\{i_1,i_2,\ldots,i_{\tau}\}\subset [N]$
{and any $D$-subset $\cR_{\cI}\subseteq [N]\setminus\cI$, the repair bandwidth
$B(\cC,\cI,\cR_{\cI})$ meets the cut-set bound described above
with equality.} Throughout this paper, such codes are also said
to have $({\tau},D)$ \emph{optimal repair} property.
\end{definition}

\subsection{Locally repairable codes}

Another important figure of merit is symbol locality~\cite{gopalan2012locality,prakash2012optimal}.

\begin{definition}\label{def_r_delta_i}
  Let $\cC$ be an $[n,k,d]_q$ linear code, and let $G$ be a generator
  matrix for it. For $j\in[n]$, the $j$-th code symbol, $c_j$, of
  $\cC$, is said to have $(r, \delta)$-locality if there exists a
  subset $S_j\subseteq [n]$ such that:
  \begin{itemize}
  \item $j\in S_j$ and $|S_j|\leq r+\delta-1$; and
  \item the minimum Hamming distance of the punctured code
    $\cC|_{S_j}$ is at least $\delta$.
  \end{itemize}
  In that case, the set $S_j$ is also called a \emph{repair set} of
  $c_j$.  The code $\cC$ is said to have information
  $(r,\delta)$-locality (denoted as $(r,\delta)_i$-locality) if there
  exists $S\subseteq [n]$ with $\rank(S)=k$ such that for each $i\in
  S$, the $i$-th code symbol has $(r, \delta)$-locality. Similarly,
  $\cC$ is said to have all symbol $(r,\delta)$-locality (denoted as
  $(r,\delta)_a$-locality) if all the code symbols have
  $(r,\delta)$-locality.
\end{definition}

In~\cite{prakash2012optimal} (and for
the case $\delta=2$, originally~\cite{gopalan2012locality}), the
following upper bound on the minimum Hamming distance of linear codes
with information $(r,\delta)$-locality is derived.

\begin{lemma}[\cite{gopalan2012locality,prakash2012optimal}] \label{lemma_bound_i}
The minimum distance, $d$, of an $[n,k,d]_q$ code with
$(r,\delta)_i$-locality, is upper bounded by
\[
  d\leq n-k+1-\left(\left\lceil\frac{k}{r}\right\rceil-1\right)(\delta-1).
\]
\end{lemma}
\begin{definition}
A code  is said to be an optimal locally repairable code (LRC) with $(r,\delta)_i$-locality (or $(r,\delta)_a$-locality) if its minimum distance $d$
attains the bound of Lemma~\ref{lemma_bound_i} with equality.
\end{definition}

We remark that in this paper, we focus only on linear codes. For
nonlinear codes with locality the reader may refer
to~\cite{forbes2014locality} and the references therein.

\section{Rack-aware distributed storage system with locality}\label{sec-rack-aware-system}
In this section, we introduce some basic model settings for a rack-aware distributed storage system with
local parity checks. In general, nodes or servers in a distributed storage system are placed on different racks, where these racks may be geographically isolated, and each rack has independent data processing capabilities.
Typically, these racks contain a small number of local parity nodes (such as those with a parity-check disk).
Additionally, to protect the data, several global parity-check racks are usually added to the system. However, existing systems are usually designed using a layered approach, for instance, employing MDS codes for encoding within the rack and then applying erasure codes at the rack level for cross rack encoding. In what follows, we will model these systems as a rack system with locality and define the repair problem for these systems. In our setting, the main idea is to consider the two-layer encoding system as a whole, that is, as a locally repairable code.

Specifically, we consider a system containing $k$ original files, which are encoded into $n$ files stored on $n$ nodes (or servers). The $n = N \times L$ nodes are divided into $N$ racks, and each rack contains $L$ nodes. In each rack, the data of the servers form a codeword with length $L$ and a minimum Hamming distance of at least $\delta$.

Due to the data processing capabilities of each rack, in each rack, we assume that all nodes within a rack have data processing capabilities and can access data from other nodes.
Denote this system as the $(n,k;L,\delta)$ rack-aware system with locality
or $(n,k;L,\delta)$-RASL. If among these $n$ nodes, $k$ of them store original information and are named as information nodes then
the system is said to be systematic. In this case, all the remaining nodes are parity checks,
including the local parity checks and cross-rack parity checks.
{ Formally, we define array codes for $(n=NL,k\leq Kr;L=r+\delta-1,\delta)$-RASL as follows.
\begin{definition}\label{def_RASL_code}
An $L\times N$ array code $\cC$ over $\F_q$ is said to be an $(N,K,k;L,\delta)$-RASL code when the code satisfies that:
\begin{itemize}
  \item {Any $K$ columns are capable of recovering the whole codeword;}
  \item {For any given $i\in [N]$, the code formed by the $i$-th columns for all codewords, i.e., the punctured code over the $i$-th column has minimum Hamming distance at least $\delta$;}
  \item {The parameter $k$ defined as $k\triangleq\log_{q}|\cC|\leq Kr$, where $r= L-\delta+1$ and the inequality follows from the well-known Singleton bound.}
\end{itemize}
\end{definition}}

\begin{remark}

In the literature, there are two different kinds of codes related to the rack-aware distributed storage system model with locality.
The first one is rack-aware regenerating codes, which are usually array codes with efficient repair schemes for failed nodes.
The other is locally repairable codes which are usually scalar codes. In this paper we consider codes for rack-aware distributed storage systems, which are related to both of these kinds. In the view of rack-aware regenerating codes, we are going to investigate rack-aware regenerating codes
such that each rack contains local parity checks. To avoid a ``three dimensional system'', we only consider the scalar case, that is,
each node is only an element of a given finite field $\F_q$ as in Definition \ref{def_RASL_code} but not a vector over $\F^{\ell}_q$ as usual. When viewed differently, codes for rack-aware distributed storage systems can be regarded as array codes formed by rearranging repair sets of scalar locally repairable codes as columns. In this viewpoint, our motivation is to build efficient repair schemes across repair sets. Thus, in what follows,
we use the notation ``$L$'', to denote the number of nodes in each rack and the length of columns for the array codes, for these two different viewpoints.
\end{remark}

\begin{figure}

\centerline{\includegraphics[width=\textwidth]{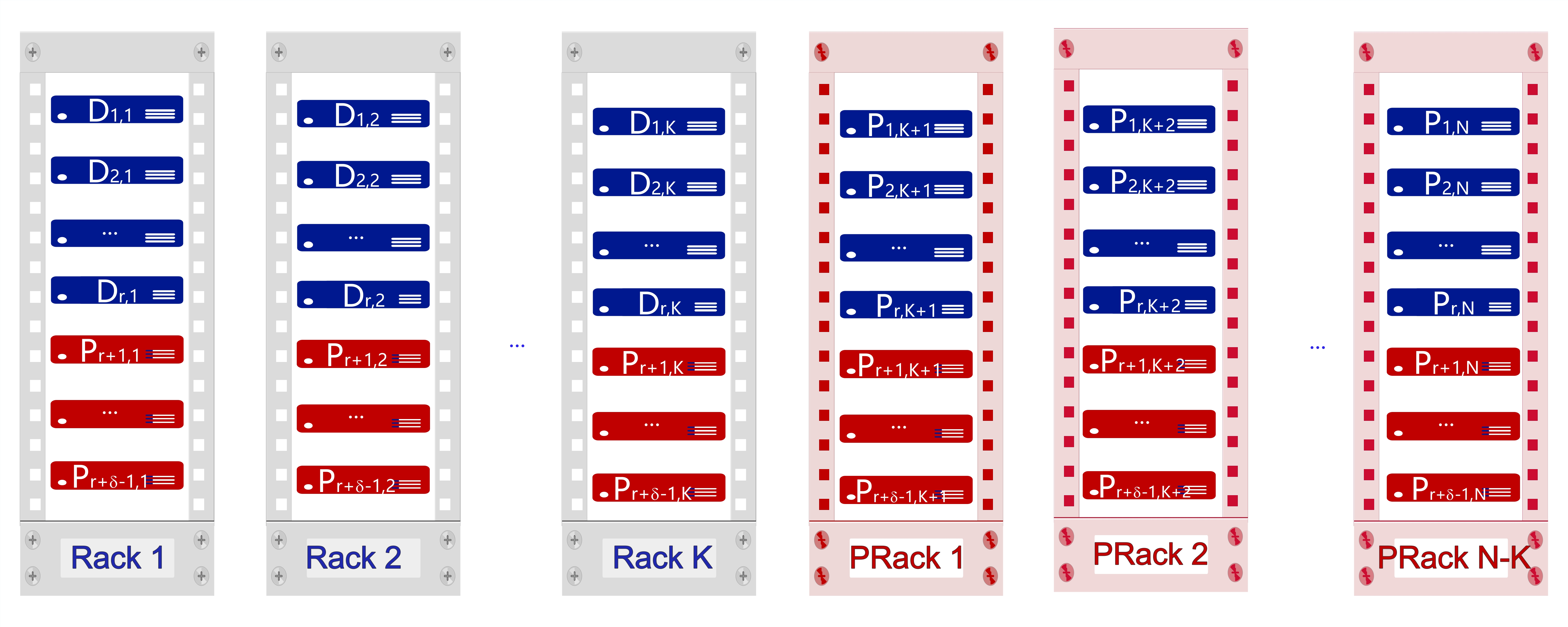}}

 \caption{The distributed storage system is organized with a racks-servers structure, where each rack consists of $r+\delta-1$ servers. Within each rack, $\delta-1$ parity checks are included to ensure data integrity. Additionally, to safeguard against rack erasures, $N-K$ rack parity checks (red columns) are stored.}
  \label{fig:RAID-like system}
\end{figure}

As an example, a systematic $(N,K,k=Kr{ ;}r+\delta-1,\delta)$-RASL is depicted in Fig. \ref{fig:RAID-like system}.

 In this system, data is distributed across several racks, each
 containing multiple nodes and  $\delta-1$ local parity
 checks, i.e., the data in each rack is erasure coded with $\delta-1$
 parity checks.  Similar to
 RAID systems that utilize disk parity checks, this storage system
 employs $N-K$ racks for across rack parity checks to handle rack erasures effectively. In this example, in total there are $n-k$ parity checks
 including $N(\delta-1)$ local parity checks and $(N-K)r$ cross-rack parity checks.

In each rack of the rack-aware system, communication between racks is more costly than
communication within a rack. Therefore, similar to rack regenerating codes \cite{hou2019rack},
we only consider the communication bandwidth across racks and disregard the inner rack bandwidth.
Since, in each rack, the data of servers form a codeword with a minimum Hamming distance of at least $\delta$, when the rack suffers $\tau\leq\delta -1$ erasures, the rack can recover the erasures locally, where we disregard the inner
rack bandwidth. Thus, we focus on racks that experience more than $\delta - 1$ server erasures.
Specifically, we consider the following two patterns of erasures:
\begin{itemize}
  \item {Rack erasures: there are $\lambda$ racks erased in the systems;}
  \item {Partial erasures: there are $\lambda$ racks that suffer { more than} $\delta-1$ nodes erasures (each).}
\end{itemize}
An intriguing challenge for these systems involves minimizing
 cross-rack bandwidth when facing erasures that exceed the
 capabilities of the local parity checks within each rack. This challenge
 serves as the motivation for the subsequent part of our
 discussion. That is, we would like to { determine} the minimum amount of data we need download from help racks to repair erasures and
the explicit construction of codes with optimal repair schemes.

\begin{remark}
{ The aforementioned rack-aware system with locality is a generalization of the rack-aware model in \cite{hou2019rack}.
For explicit constructions of regenerating codes for rack-aware system the reader may refer to \cite{hu2017optimal,hou2019rack,chen2020explicit,wang2023rack,zhang2023vertical}. }
This extension is primarily driven by the practical observation that modern storage systems incorporate both parity checks and controllers with data processing capabilities within each rack. In Fig. \ref{fig:RASL}, we compare the rack-aware model and rack-aware model with locality for $n=18$, $k=10$, and $r=5$, where we set $\lambda=1$ for both rack erasure and partial erasure cases.
\begin{figure}
\begin{minipage}{0.32\linewidth}
\centerline{\includegraphics[width=\textwidth]{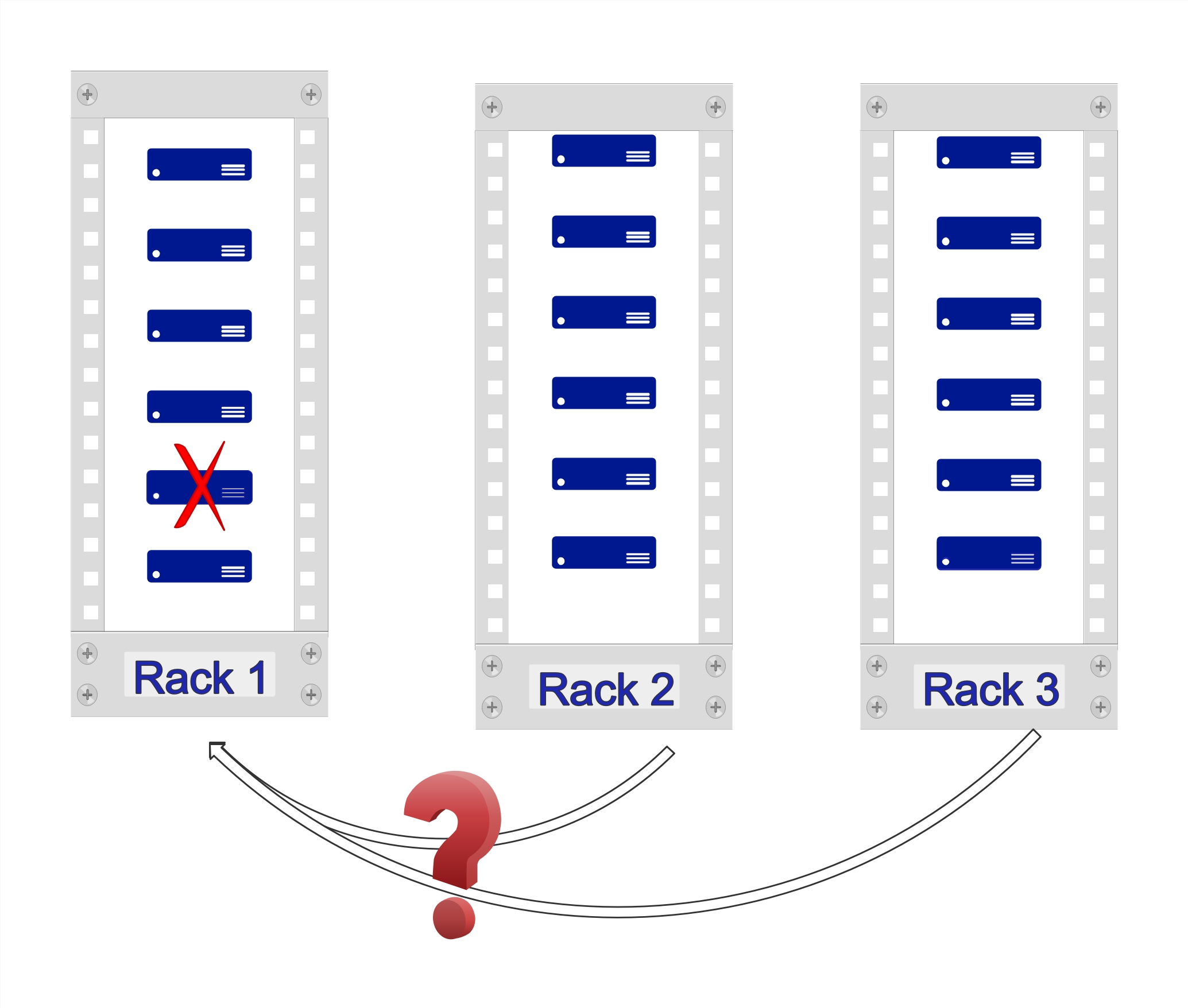}}
\end{minipage}
\begin{minipage}{0.32\linewidth}
\centerline{\includegraphics[width=\textwidth]{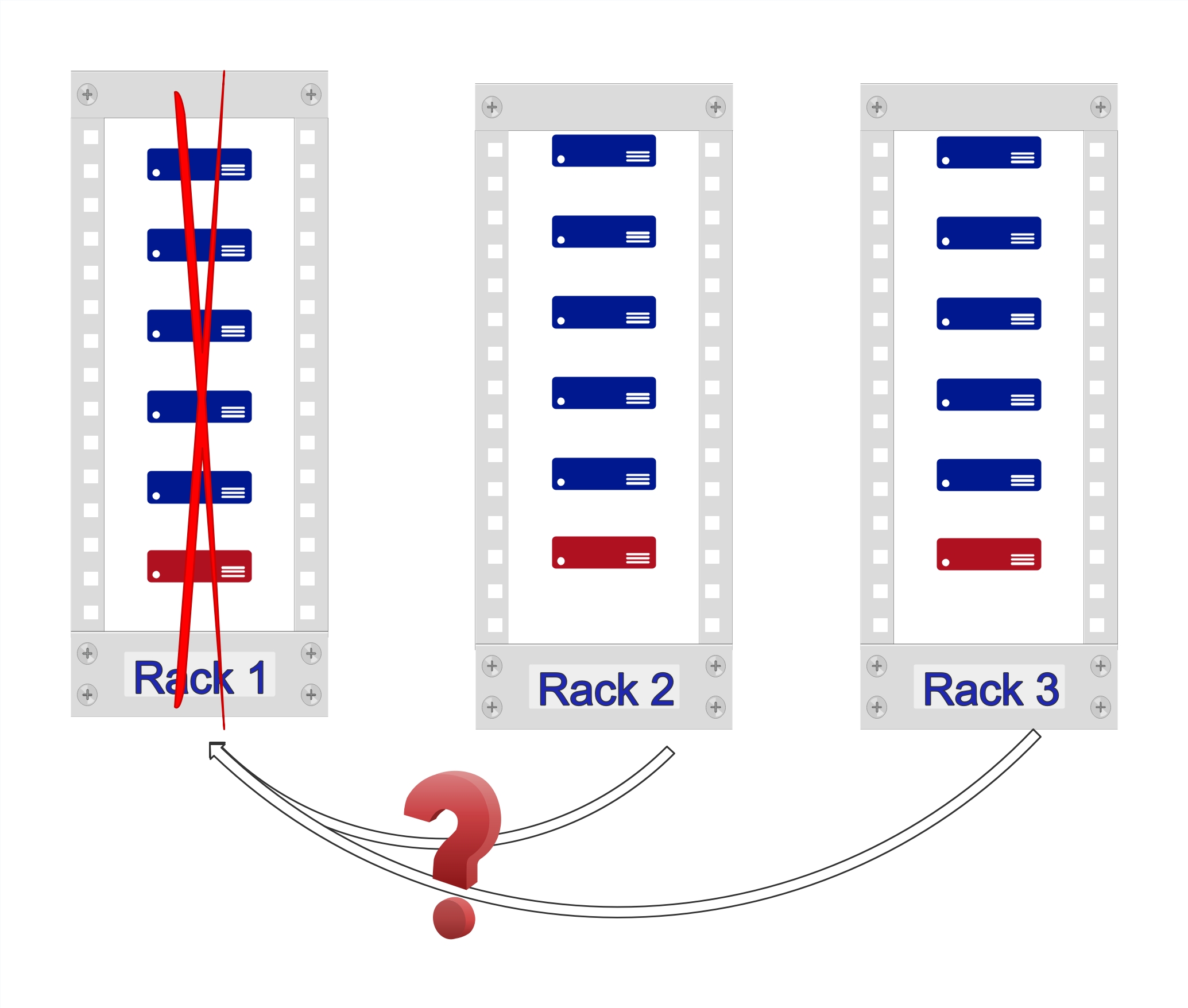}}
\end{minipage}
\begin{minipage}{0.32\linewidth}
\centerline{\includegraphics[width=\textwidth]{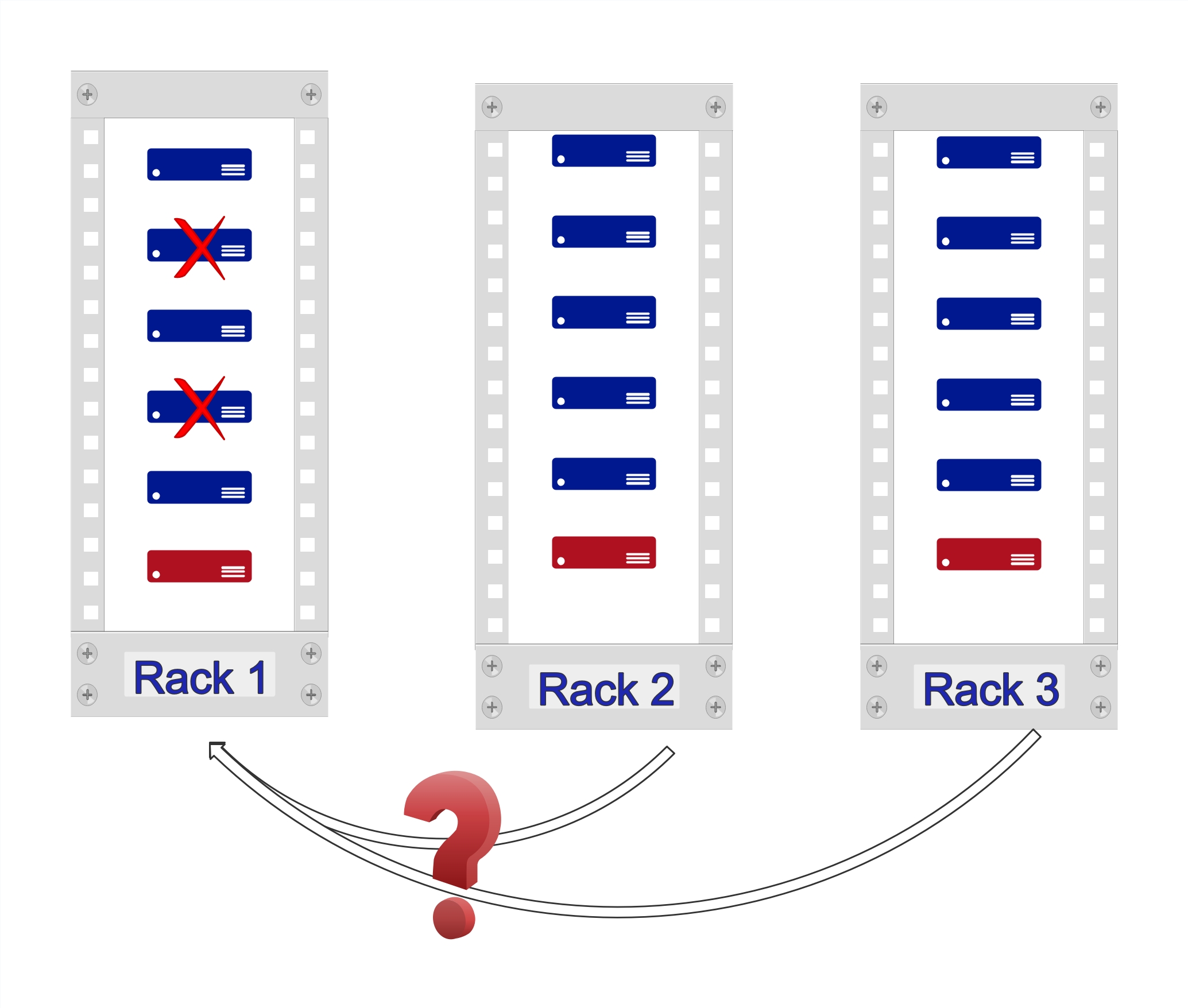}}
\end{minipage}
\caption{As a comparison, setting $n=18$, $k=10$, the first figure shows the rack-aware model with one erasure, the second one is corresponding to the $(18,10;6,2)$-RASL for one rack erasure and the last is for the $(18,10;6,2)$-RASL with one partial erasure, where the red nodes refer to the local parity checks. }
  \label{fig:RASL}
\end{figure}

\end{remark}

\begin{remark}
The term ``locality'' is derived from the condition that, for $(n, k)$ locally repairable codes with $(r, \delta)_a$-locality, if the repair sets $S_1,S_2,\cdots, S_N$ constitute a partition of $[n]$ and each set has a uniform size, i.e., $|S_i|=r+\delta-1$ for $1\leq i\leq N$, then each repair set can be arranged as a rack to construct the desired code for a rack-aware system. Consequently, when the repair sets form a partition, the repair problems for the rack-aware system with locality are equivalent to the repair problems of locally repairable codes, assuming that we disregard the bandwidth within the rack or the repair set. Therefore, in the subsequent discussion, we will use these notations interchangeably. As in Figure \ref{fig:RASL}, we may apply a $(18,10)$ locally repairable code with $(5,2)$-locality in the
$(18,10;6,2)$-RASL, when there are three disjoint repair sets with size $6$ corresponding to three racks.
\end{remark}

\begin{remark}
To simplify the system setting, we focus on the regular case where each rack contains an equal number of servers.
\end{remark}

\section{Tamo-Barg codes and redundant residue codes}\label{sec_TB_CRT}

In \cite{tamo2014family}, Tamo and Barg proposed constructions of
locally repairable codes. Among them, one of the constructions yields
codes now named Tamo-Barg codes, and another is via the Chinese
Remainder Theorem, namely, redundant residue codes. Tamo-Barg codes
are in fact a special case of locally repairable codes via the Chinese
Remainder Theorem~\cite{tamo2014family}. However, in general, the
locality of parity-check symbols for locally repairable codes based on
the Chinese Remainder Theorem is not well understood. In this section,
we review the construction of locally repairable codes using redundant
residue codes and prove that it can, in fact, explain Tamo-Barg
codes. We begin with the Chinese Remainder Theorem for
polynomials over finite fields.

\begin{lemma}(\cite{pei1996chinese})\label{lemma_CRT}
Let $h_1(x),\dots,h_{t}(x)\in \F_q[x]$ be pairwise co-prime
polynomials. Then for any $t$ polynomials $m_1(x),\dots,$ $m_t(x)\in
\F_q[x]$, there exists a unique polynomial $f(x) \in \F_q[x]$ of
degree less than $\sum_{i=1}^{t}\deg(h_i(x))$ such that
\begin{equation*}
f(x)\equiv m_i(x)\pmod{h_i(x)}\quad \text{ for all } i\in[t].
\end{equation*}
\end{lemma}

\begin{construction}[\cite{tamo2014family}]
  \label{cons_RRC}
  Let $h(x)\in \F_q[x]$ and denote $\deg(h(x))=w$. For $y\in \F_q$,
  define $\roots(y)\eqdef\{x\in \F_q ~:~ h(x)=y\}$ and
  $t_y\eqdef|\roots(y)|$.  Assume $m_1\leq k$ and $r_i$, for all
  $i\in[m_1]$, are positive integers such that $\sum_{i=1}^{m_1}r_i=
  k$. We further assume that there exist two disjoint subsets of
  $\F_q$, $\{y_i\}_{i=1}^{m_1}$ and $\{y_i\}_{i=m_1+1}^{m_1+m_2}$,
  satisfying $t_{y_i}>r_i$ for all $i\in[m_1]$, and $m_2$ is a
  non-negative integer.

Denote $m\eqdef m_1+m_2$, and $\roots(y_i)=\{\beta_{i,1}, \beta_{i,2},
\ldots, \beta_{i,t_{y_i}}\}$ for all $i\in[m]$.  Let
$\bma=(a_{1,1},a_{1,2},\dots,a_{1,r_1},a_{2,1},\dots,$ $a_{m_1,r_{m_1}})\in
\F_q^k$ be the information vector.  Define $f_{\bma,i}$ as the
polynomial with degree less than $r_i$ such that
$f_{\bma,i}(\beta_{i,j})=a_{i,j}$ for all $i\in[m_1]$ and
$j\in[r_i]$. Let $F_{\bma}(x) \in \F_q[x]$ be a polynomial with degree
less than $m_1w$ satisfying
\begin{equation}\label{eqn_F_a}
  F_{\bma}(x)\equiv f_{\bma,i}(x)\pmod{h(x)-y_i}\quad\text{ for all } i\in[m_1].
\end{equation}
Then the code we construct is
\[\cC=\{C_{\bma}=(F_{\bma}(\beta_{1,1}),F_{\bma}(\beta_{1,2}),\dots,F_{\bma}(\beta_{1,t_{y_1}}) \dots, F_{\bma}(\beta_{m,t_{y_m}})) ~:~ \bma\in \F^k_q\}.\]
\end{construction}

\begin{remark}
By Lemma~\ref{lemma_CRT}, the fact that $\gcd(h(x)-y_i,h(x)-y_j)=1$ for distinct $i,j\in[m_1]$ implies that ${F}_{\bma}(x)$ and $C_{\bma}$ are well defined.
\end{remark}

According to \eqref{eqn_F_a}, determining the locality of information
symbols in $\cC$ is straightforward, as discussed in~\cite[Theorem
  5.3]{tamo2014family}. In the following, we present a lemma that is
useful for determining the locality of global parity-check symbols.

\begin{lemma}\label{lemma_F_a}
  Consider the setting of Construction~\ref{cons_RRC}, and let $0 \le
  r \le w$ be an integer.  Suppose that there exist $m_1$ distinct
  constants $y_1,y_2,\dots,y_{m_1}\in \F_q$ such that
\[
F_{\bma}(x)\pmod{h(x)-y_i} = f_{\bma, i}(x)=\sum_{j=0}^{r-1}e_{i,j}x^{j} \quad \text{ for all } i\in[m_1].
\]
Then for any $y\in\F_q$,
\[
F_{\bma}(x)\pmod{h(x)-y} = \sum_{j=0}^{r-1}H_{\bma,j}(y)x^{j},
\]
where $H_{\bma,j}(x)$ is a polynomial satisfying $\deg(H_{\bma,j}(x))\leq m_1-1$
and
\[
H_{\bma,j}(y_i)=e_{i,j}\text{ for all $i\in[m_1]$ and $0\leq j\leq r-1$.}
\]
\end{lemma}

\begin{IEEEproof}
  Let us divide $F_{\bm a}(x)$ by $h(x)$ to obtain a quotient $U_1(x)$
  and a remainder $R_1(x)$,
  \[F_{\bm a}(x)=h(x)U_1(x)+R_1(x),\]
  with $\deg(R_1(x))<w=\deg(h(x))$, $\deg(U_1(x))=\deg(F_{\bm
    a}(x))-w< (m_1-1)w$ and $U_1(x),\,R_1(x)\in \F_q[x]$. We also have
  \[F_{\bma}(x)=(h(x)-y)U_1(x)+yU_1(x)+R_1(x).\]
  Similarly, we can write $U_1(x)$ as
  \[U_1(x)=h(x)U_2(x)+R_2(x).\]
  If $\deg(U_1(x))\geq w$, then $\deg(R_2(x))<w$ and $\deg(U_2(x)) <
  (m_1-2)w$. However, if $\deg(U_1(x))< w$, we set $U_2(x)=0$ and
  $R_2(x)=U_1(x)$. Then
\begin{equation*}
\begin{split}
F_{\bma}(x)=&(h(x)-y)U_1(x)+y(h(x)U_2(x)+R_2(x))+R_1(x)\\
=&(h(x)-y)(U_1(x)+yU_2(x))+y^2U_2(x)+yR_2(x)+R_1(x).
\end{split}
\end{equation*}
Repeating this procedure $m_1-1$ times, we conclude that
\[U_{m_1-2}(x)=h(x)U_{m_1-1}(x)+R_{m_1-1}(x),\]
with
$\deg(U_{m_1-1}(x))<w$, $\deg(R_{m_1-1}(x))<w$,  and
\begin{align*}
 F_{\bma}(x)&=(h(x)-y)(U_1(x)+yU_2(x))+y^2U_2(x)+yR_2(x)+R_1(x)\\
 &=(h(x)-y)(U_1(x)+yU_2(x){ +\dots+}y^{m_1-2}U_{m_1-1}(x)) \\
 &\quad +y^{m_1-1}U_{m_1-1}(x)+y^{m_1-2}R_{m_1-1}(x)+\dots+yR_2(x)+R_1(x).
\end{align*}
Note that $\deg(U_{m_1-1}(x))<\deg(h(x))=w$ and $\deg(R_i(x))<w$
for all $i\in[m_1-1]$. Thus,
\begin{equation}\label{eqn_R_i}
F_\bma(x)\pmod{ h(x)-y} = y^{m_1-1}U_{m_1-1}(x)+y^{m_1-2}R_{m_1-1}(x)+\dots+yR_2(x)+R_1(x)\eqdef H_{\bma}(x,y).
\end{equation}

Let us now rewrite $H_\bma(x,y)$ as
\begin{equation}\label{eqn_H_x_c}
H_{\bm a}(x,y)=H_{\bma,w-1}(y)x^{w-1}+H_{\bma,w-2}(y)x^{w-2}+\dots+H_{\bma,0}(y)x^{0},
\end{equation}
where $H_{\bma,j}(y)=\sum_{i=0}^{m_1-1}{h_{i,j}}y^i$ and $h_{i,j}$ is
the coefficient of $x^{j}$ for $R_{i+1}(x)$ if $0\leq i\leq m_1-2$,
and the coefficient of $x^{j}$ for $U_{m_1-1}(x)$ if $i=m_1-1$. Thus,
$H_{\bma ,j}(y)$ for $0\leq j\leq w-1$ can be regarded as polynomials
in $y$. According to the assumption, there exist $m_1$ constants
$y_1,y_2,\dots,y_{m_1}$ such that ${H_{\bma}}(x,y_i)$ for $i\in[m_1]$
are polynomials with degree less than $r$ in $x$. Hence,
\begin{equation*}
H_{\bma, j}(y_i)=0\qquad \text{ for $r\leq j\leq w-1$, and $i\in[m_1]$},
\end{equation*}
which means $H_{\bma,j}(y)\equiv0$ for $r\leq j\leq w-1$ since
$\deg({H_{\bma,j}}(y))<m_1$ for ${r\leq j\leq w-1}$,
and then
\begin{equation*}
F_{\bma}(x)\pmod{h(x)-y}=\sum_{j=0}^{r-1}H_{\bma,j}(y)x^{j}.
\end{equation*}
The first claimed result follows directly from \eqref{eqn_R_i} and
\eqref{eqn_H_x_c}.  For the second part, since $\deg(H_{\bm
  a,j}(x))\leq m_1-1$ for $0\leq j\leq r-1$, the rest of the lemma
follows from the fact that there exists a unique polynomial $H_{\bma,
  j}(y)$ for $0\leq j\leq r-1$ such that
\begin{equation*}
H_{\bma, j}(y_i)=e_{i,j} \qquad\text{ for all $i\in[m_1]$,}
\end{equation*}
where
$f_{\bma,i}(x)=\sum_{j=0}^{r-1}e_{i,j}x^j= F_{\bma}(x)\pmod {h(x)-y_i}$.
This completes the proof.
\end{IEEEproof}

\begin{corollary}\label{coro_RRC_gen}
  In the setting of Construction~\ref{cons_RRC}, let
  $\Gamma=\bigcup_{i\in[m]}\roots(y_i)\subseteq \F_q$. If
  $\abs{\roots(y_i)}>r$ for any $i\in[m]$, then the code
  constructed by Construction \ref{cons_RRC} is a locally repairable
  code with all symbol $(r,\delta)$-locality, where
  $\delta=\min\{\abs{\roots(y_i)}+1-r ~:~ i\in[m]\}$.
\end{corollary}

Corollary \ref{coro_RRC_gen} follows directly from Lemma
\ref{lemma_F_a}, and we omit its proof. In general, the code may not
be optimal for the simple reason that there may not be enough roots in
$\F_q$ for some $h(x)-y_i$, where $i\in[m]$, to serve as evaluation
points. Therefore, to attain optimality, we consider the case
where $\F_q$ contains the splitting field of $h(x)-y_i$, for all $i\in[m]$.

\begin{definition}
  In the setting of Construction~\ref{cons_RRC}, let
  $\Gamma=\bigcup_{i\in[m]}\roots(y_i)\subseteq \F_q$. If
  $\abs{\roots(y_i)}=\deg(h(x))=w$ for all $i\in[m]$, then the
  polynomial $h(x)$ is said to be a \emph{good polynomial} over
  $\Gamma$.
\end{definition}

For more details about good polynomials the readers may refer to
\cite{tamo2014family} and \cite{liu2017new}.

\begin{corollary}[Tamo-Barg codes,\cite{tamo2014family}]\label{coro_TB_code}
Consider the setting of Construction~\ref{cons_RRC} and let $h(x)$ be
a good polynomial over $\Gamma=\bigcup_{i\in[m]}\roots(y_i)\subseteq
\F_q$.  If $r_i=r<w$ and $k=rm_1$, then the resulting code is an
optimal $[n=mw,m_1r,(m_2+1)w-r+1]_q$ locally repairable code with all
symbol $(r,w-r+1)$-locality (where optimality is with respect to the bound in Lemma~\ref{lemma_bound_i}).
\end{corollary}

Corollary \ref{coro_TB_code} follows directly from
Lemma~\ref{lemma_F_a} and Corollary~\ref{coro_RRC_gen}, which also can
be directly derived from \cite[Construction 1]{tamo2014family}. We
{ omit} its proof.  This result was first introduced in
\cite{tamo2014family}.


\section{Repairing Tamo-Barg codes: rack erasures}\label{sec-TBcode-RRC}
In this section, we consider the repair problem for Tamo-Barg
codes for the rack erasure case, where the repair sets are arranged as racks. We begin with an array form of the Tamo-Barg code,
where each repair set is arranged as a column in the array.

\begin{construction}\label{cons_TB_array}
Let $h(x)\in\F_q[x]$ be a good polynomial over
$\Gamma=\bigcup_{i\in[m]}\roots(y_i)\subseteq \F_q$ with
$\deg(h(x))=r+\delta-1$, and let
$\bma=(a_{1,1},a_{1,2},\dots,a_{1,r},a_{2,1},\dots a_{m_1,r})\in
\F_q^k$ be the information vector, where $k=rm_1$.  Define $f_{\bm
  a,i}(x)$ as the polynomial with degree less than $r$ such that
$f_{\bma,i}(\beta_{i,j})=a_{i,j}$ for all $i\in[m_1]$ and $j\in[r]$,
where we assume that $\roots(y_i)=\{\beta_{i,1}, \beta_{i,2}, \dots,
\beta_{i,r+\delta-1}\}$ for all $i\in[m_1]$. Then for any $\bma\in
\F_q^k$ we can find a polynomial $F_{\bma}(x) \in \F_q[x]$ with degree
less than $m_1(r+\delta-1)$ satisfying
\[
F_\bma(x)\equiv f_{\bma,i}(x)\pmod{h(x)-y_i}\qquad\text{for all $i\in[m_1]$}.
\]
Construct an array code as follows
\begin{equation*}
\small
\cA\eqdef\bracenv{A_{\bma}=\parenv{\bA_1,\bA_2,\dots,\bA_{m}}=\parenv{
\begin{matrix}
&F_\bma(\beta_{1,1})&F_\bma(\beta_{2,1})&\dots &F_\bma(\beta_{m,1})\\
&F_\bma(\beta_{1,2})&F_\bma(\beta_{2,2})&\dots &F_\bma(\beta_{m,2})\\
&\vdots &\vdots& &\vdots\\
&F_\bma(\beta_{1,r+\delta-1})&F_\bma(\beta_{2,r+\delta-1})&\dots &F_\bma(\beta_{m,r+\delta-1})\\
\end{matrix}
  }
  ~:~
  \bma\in \F^k_q},
\end{equation*}
where we define $m\eqdef m_1+m_2$.
\end{construction}

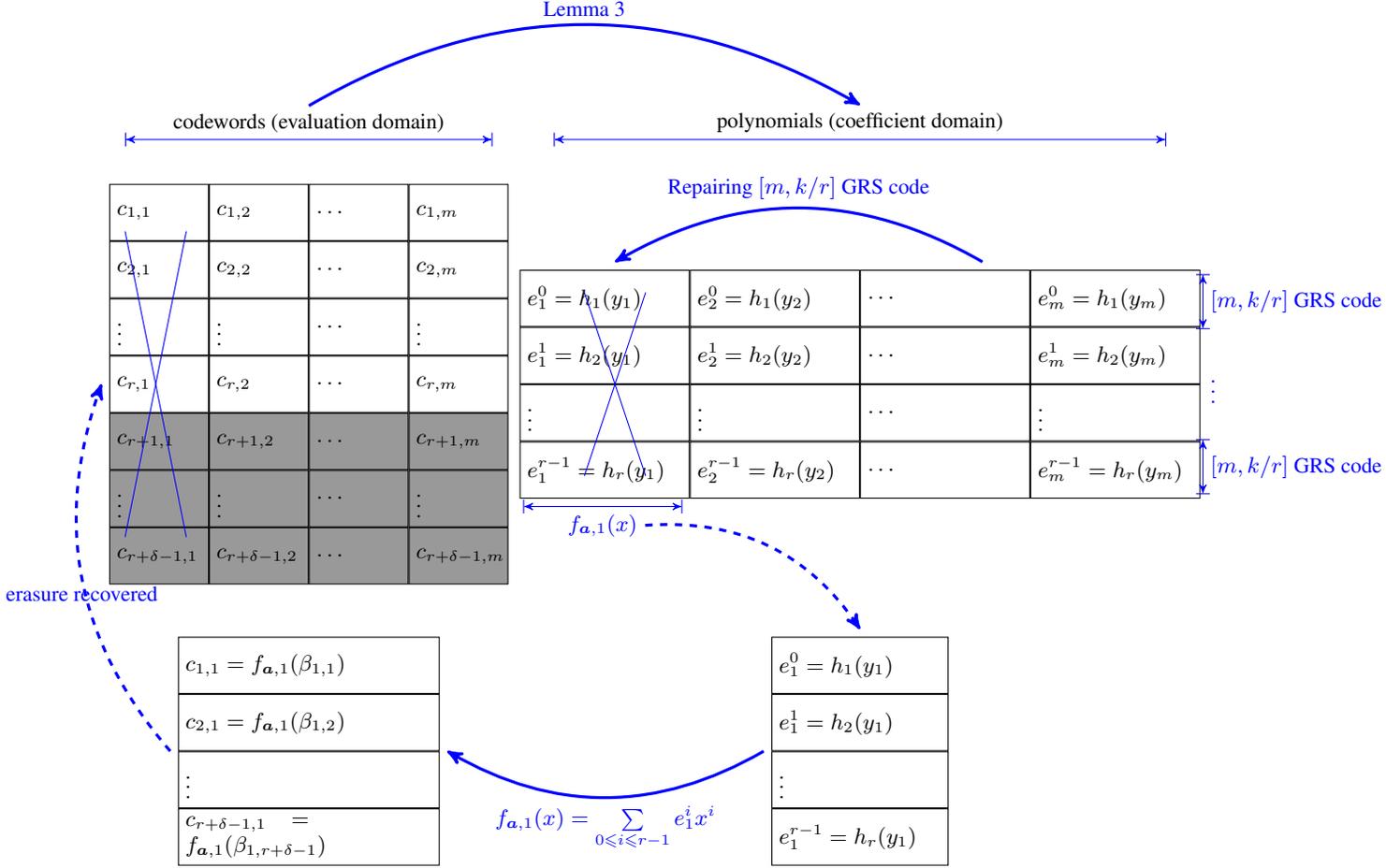
\begin{figure}[tb]
  \centering
\scalefont{0.87}
\begin{tikzpicture}[scale=0.87]
   \draw [|<->|, blue] (-7,4) -- (-1,4);
   \node (code)  at (-4,4) [above,black]{codewords (evaluation domain)};
    \matrix [nodes={text width=12mm,draw},nodes={minimum size=8mm}](Ori_code) at (-4,0)
    {\node{$c_{1,1}$};&\node{$c_{1,2}$};&\node{$\cdots$};
    &\node{$c_{1,m}$};\\
    \node{$c_{2,1}$};&\node{$c_{2,2}$};&\node{$\cdots$};
    &\node{$c_{2,m}$};\\
    \node{$\vdots$};&\node{$\vdots$};&\node{$\cdots$};&\node{$\vdots$};\\
    \node{$c_{r,1}$};&\node{$c_{r,2}$};&\node{$\cdots$};&\node{$c_{r,m}$};\\
    \node[fill=light_gray]{$c_{r+1,1}$};&\node[fill=light_gray]{$c_{r+1,2}$};&\node[fill=light_gray]{$\cdots$};&\node[fill=light_gray]{$c_{r+1,m}$};\\
    \node[fill=light_gray]{$\vdots$};&\node[fill=light_gray]{$\vdots$};&\node[fill=light_gray]{$\cdots$};&\node[fill=light_gray]{$\vdots$};\\
    \node[fill=light_gray]{$c_{r+\delta-1,1}$};&\node[fill=light_gray]{$c_{r+\delta-1,2}$};&\node[fill=light_gray]{$\cdots$};&\node[fill=light_gray]{$c_{r+\delta-1,m}$};\\
    };

    \draw [|<->|, blue] (10,4) -- (0,4);
    \node (coef) at (5,4) [above,black]{polynomials (coefficient domain)};

    \matrix [nodes={text width=22mm,draw},nodes={minimum size=8mm}] at (5,0)
    {\node{$e^{0}_1=h_1(y_1)$};&\node{$e^0_2=h_1(y_2)$};&\node{$\cdots$};
    &\node{$e^0_m=h_1(y_m)$};\\
    \node{$e^1_1=h_2(y_1)$};&\node{$e^1_2=h_2(y_2)$};&\node{$\cdots$};
    &\node{$e^1_m=h_2(y_m)$};\\
    \node{$\vdots$};&\node{$\vdots$};&\node{$\cdots$};&\node{$\vdots$};\\
    \node{$e^{r-1}_1=h_{r}(y_1)$};&\node{$e^{r-1}_2=h_{r}(y_2)$};
    &\node{$\cdots$};&\node{$e^{r-1}_m=h_{r}(y_m)$};\\
    };
    \draw [blue] (-7,-2.5)--(-6,2.5);
    \draw [blue] (-7,2.5)--(-6,-2.5);
    \draw [blue] (0.5,-1.5)--(1.5,1.5);
    \draw [blue] (0.5,1.5)--(1.5,-1.5);

    \draw [|<->|, blue] (10.6,1.8) -- (10.6,0.9) node at (10.6,1.35) [right, blue]{$[m,k/r]$ GRS code};
    \node at (10.6,0) [right, blue]{$\vdots$};
    \draw [|<->|, blue] (10.6,-1.8) -- (10.6,-0.9) node at (10.6,-1.35) [right, blue]{$[m,k/r]$ GRS code};
    \draw [|<->|, blue] (-0.5,-2) -- (2.1,-2);
    \node (f1) at (0.8,-2) [below, blue]{$f_{\bm a,1}(x)$};

    \matrix [nodes={text width=23mm,draw},nodes={minimum size=8mm}](mat_coef) at (5,-6)
    {\node{$e^0_1=h_1(y_1)$};\\
    \node{$e^1_1=h_2(y_1)$};\\
    \node{$\vdots$};\\
    \node{$e^{r-1}_1=h_{r}(y_1)$};\\};

\matrix [nodes={text width=35mm,draw},nodes={minimum size=8mm}](mat_code) at (-4,-6)
    {\node{$c_{1,1}=f_{\bm a, 1}(\beta_{1,1})$};\\
    \node{$c_{2,1}=f_{\bm a, 1}(\beta_{1,2})$};\\
    \node{$\vdots$};\\
    \node{$c_{r+\delta-1,1}=f_{\bm a, 1}(\beta_{1,r+\delta-1})$};\\
    };

    \path[blue,->,very thick,bend right] (7,2) edge node[above] {Repairing $[m,k/r]$ GRS code} (1,2);
    \path[blue,->,very thick,bend left](code.north)edge node[above] {Lemma \ref{lemma_F_a}} (coef.north);
	\path[blue,->,very thick,bend left,dashed](f1.east)edge node[above] {} (mat_coef.north);
    \path[blue,->,very thick,bend left](mat_coef.west)edge node[below] {$f_{\bm a, 1}(x)=\sum\limits_{0\leq i\leq r-1}e^i_1x^i$} (mat_code.east);
    \path[blue,->,very thick,bend left,dashed](mat_code.west)edge node[below] {erasure recovered} (Ori_code.west);
\end{tikzpicture}

  \caption{Repairing Tamo-Barg codes. The repair problem of the first column is reduced to the repair problem for the corresponding polynomial remainder. Secondly, by Lemma \ref{lemma_F_a}, this is equivalent to the repair problem of certain codewords within a Reed-Solomon code, where each of them suffers one erasure. Finally, the repairing of these codewords ensures the recovery of the erased column.}
  \label{fig:repairing}
\end{figure}

Now, we are going to present repair schemes for the array-form Tamo-Barg codes, whose main idea
is explained in Fig. \ref{fig:repairing}.

\begin{theorem}\label{theorem_repair_TB}
  Consider the setting of Construction~\ref{cons_TB_array}. Let $q_0$
  be a prime power, and $\F_q=\F_{q_0}(y_1,y_2,\dots,y_{m})$,
  $\F_{q_i}=\F_{q_0}(y_1,\dots,y_{i-1},y_{i+1},\dots,y_{m})$. Define $w_i\eqdef [\F_q:\F_{q_i}]$ for each $i\in[m]$.
  \begin{itemize}
  \item[I.] Define $w^*_i\triangleq\frac{r}{\gcd(w_i,r)}$.  If
    $w^*_i\leq w_i$ and $m_2\geq \max\{2,w^*_i\}$, then we can recover
    $\bA_i$ by downloading { $(w^*_i+m_1-1)r$ elements  of $\F_{q_i}$, i.e.,} $\frac{(w^*_i+m_1-1)r}{w^*_i}$ symbols { in $\F_q$} from
    any other $w^*_i+m_1-1$ { racks} (columns), where each symbol is an element of $\F_{q}$.
  \item[II.] Assume $w^*_i$ is a positive integer. If $w^*_i|w_i$ and $m_2\geq \max\{2,w^*_i\}$, then we
    can recover $\bA_i$ by downloading { $(w^*_i+m_1-1)r$ elements of $\F_{q_i}$, i.e.,} $\frac{(w^*_i+m_1-1)r}{w^*_i}$
    symbols { in $\F_q$} from any other $w^*_i+m_1-1$ { racks} (columns), where each symbol denotes an element of $\F_q$.
\end{itemize}
\end{theorem}

\begin{IEEEproof}
  We begin by proving claim I. To recover
  \[\bA_i=(F_\bma(\beta_{i,1}),F_\bma(\beta_{i,2}),\dots,F_\bma(\beta_{i,r+\delta-1}))^\top
  =({f_{\bma,i}(\beta_{i,1}),f_{\bma,i}(\beta_{i,2}),\dots,f_{\bma,i}(\beta_{i,r+\delta-1})})^\top,\]
  it suffices that we recover $f_{\bma,i}(x)\equiv
  F_\bma(x)\pmod{h(x)-y_i}$.  By Lemma \ref{lemma_F_a}, we only need
  to figure out $H_{\bm a,j}(y_i)$ for $0\leq j\leq r-1$. Note that
  for $t\in[m]\setminus\{i\}$ we have
\[\bA_t=(F_\bma(\beta_{t,1}),F_\bma(\beta_{t,2}),\dots,F_\bma(\beta_{t,r+\delta-1}))^\top=(f_{\bma, t}(\beta_{t,1}),f_{\bma,t}(\beta_{t,2}),\dots,f_{\bma, t}(\beta_{t,r+\delta-1}))^\top\]
and $f_{\bma,t}(x)=\sum_{j=0}^{r-1}H_{\bma,j}(y_t)x^j$. Thus, based on
$\bA_t$, we can calculate $H_{\bma,j}(y_t)$ for $t\in[m]\setminus\{i\}$ and $0\leq j\leq r-1$.

We observe that $w^*_i| r$, which implies that we can divide $H_{\bm
  a,j}(y_t)$ for $0\leq j\leq r-1$ into $r/w^*_i$ vectors of length
$w^*_i$, say
\[
\bH_{\bma}(y_t,\tau)\triangleq \parenv{H_{\bma,\tau w^*_i}(y_t),H_{\bma,\tau w^*_i+1}(y_t),\dots,H_{\bma,(\tau+1) w^*_i-1}(y_t)},
\]
for $0\leq \tau\leq r/w^*_i-1$.
Let $\F=\F_q(\beta)$ such that $[\F:\F_q]=w^*_i$.
{Then we} have
$[\F:\F_{q_i}]=w_i^*w_i$ and $\{\beta^ty^j_i ~:~ 0\leq t\leq w^*_i-1,\, 0\leq j\leq w_i-1\}$
is a {basis} of $\F$ over $\F_{q_i}$. Let $\Psi$ be the bijection from $\F^{w^*_i}_q$ to $\F$
\[
\Psi(V)=\sum_{j=0}^{w^*_i-1}{v_j}\beta^j,
\]
where $V=(v_0,v_1,\dots,v_{w^*_i-1})\in \F^{w^*_i}_q$.
Now, for any $0\leq \tau\leq r/w^*_i-1${, we} may regard
$$(\Psi(\bH_\bma(y_1,\tau)),\Psi(\bH_\bma(y_2,\tau)),\dots,\Psi(\bH_\bma(y_{m_1+m_2},\tau)))$$
 as a codeword of a GRS code $\cC_1$
with parameters $[m_1+m_2,m_1,m_2+1]_{q^{w^*_i}}$, since it corresponds to
evaluations of a polynomial
\[H_\bma^{(\tau)}(x)=\sum_{j=0}^{w^*_i-1}\beta^{j}H_{\bma,\tau w^*+j}(x)\in \F[x],\]
with $\deg(H_\bma^{(\tau)}(x))<m_1$. Now the repair problem of
$H_{\bma, j}(y_i)$ (i.e., $\bH_\bma(y_i,\tau)$) is exactly the repair
problem of a GRS code.  For completeness, we include some known methods
for repairing GRS codes
\cite{guruswami2017repairing,tamo2017optimal,tamo2018repair}.

Let $\Theta=\{y_{j_t} ~:~ 1\leq t\leq w^*_i+m_1-1\}$ be any
$(w^*_i+m_1-1)$-subset of $\Gamma_i=\{y_j ~:~
j\in[m]\}\setminus\{y_i\}$. Define
\begin{equation}\label{eqn_g_theta}
g_\Theta(x)\triangleq \prod_{\theta\in \Gamma_i\setminus \Theta}(x-\theta).
\end{equation}
By \eqref{eqn_g_theta}, we have $\deg(x^ug_\Theta(x))<m_2$ for $0\leq
u\leq w^*_i-1$. Note that the dual code of $\cC_1$ is also a GRS code,
with parameters $[m_1+m_2,m_2,m_1+1]_{q^{w^*_i}}$. Thus, there exists
a vector $(v_1,v_2,\dots,v_{m_1+m_2})\in (\F^*)^{m_1+m_2}$ such that
\begin{equation*}
(v_1y_1^ug_\Theta(y_1),v_2y_2^ug_\Theta(y_2),\dots,v_{m_1+m_2}y^u_{m_1+m_2}g_\Theta(y_{m_1+m_2}))\in \cC^{\bot}_1\quad\text{for all }0\leq u\leq w^*_i-1,
\end{equation*}
i.e.,
\begin{equation}\label{eqn_dual}
\sum_{j=1}^{m_1+m_2}v_jy_j^ug_\Theta(y_j)\Psi(\bH_\bma(y_j,\tau))=0\quad\text{for all $0\leq u\leq w^*_i-1$ and $0\leq \tau\leq \frac{r}{w^*_i}-1$.}
\end{equation}
For $\{y^u_i ~:~ 0\leq u\leq w^*_i-1\}$, from Lemma 1 in
\cite{tamo2018repair}, we can find a set $\{\gamma_t ~:~ 0\leq t\leq
w_i-1\}$ such that $\{y_i^u\gamma_t ~:~ 0\leq u \leq w^*_i-1,\,0\leq
t\leq w_i-1\}$, just as $\{\beta^ty^j_i ~:~ 0\leq t\leq w^*_i-1,\,
0\leq j\leq w_i-1\}$, is a {basis} of $\F$ over $\F_{q_i}$.  By
\eqref{eqn_dual}, for $0\leq u\leq w^*_i-1$, $0\leq \tau\leq
r/w^*_i-1$, and $0\leq t\leq w_i-1$,
\[
\begin{split}
\tr_{\F/\F_{q_i}}\parenv{\gamma_t v_iy_i^ug_\Theta(y_i)\Psi(\bH_\bma(y_i,\tau))}=&-\sum_{\substack{j\in[m_1+m_2]\\ j\ne i}}
\tr_{\F/\F_{q_i}}\parenv{\gamma_t v_jy_j^ug_\Theta(y_j)\Psi(\bH_\bma(y_j,\tau))}\\
=&-\sum_{\substack{j\in[m_1+m_2]\\ j\ne i}}y_j^ug_\Theta(y_j)\tr_{\F/\F_{q_i}}\parenv{\gamma_t v_j
\Psi(\bH_\bma(y_j,\tau))}\\
=&-\sum_{\substack{j\in[m_1+m_2]\\ j\ne i,\,y_j\in \Theta}}y_j^ug_\Theta(y_j)\tr_{\F/\F_{q_i}}\parenv{\gamma_t v_j\Psi(\bH_\bma(y_j,\tau))},\\
\end{split}
\]
where the last equality holds by \eqref{eqn_g_theta}.  As already
mentioned, $\{y_i^u\gamma_t ~:~ 0\leq u \leq w^*_i-1,\,0\leq t\leq
w_i-1\}$ is a basis of $\F$ over $\F_{q_i}$, so
$\{\tr_{\F/\F_{q_i}}\parenv{\gamma_t
  v_iy_i^ug_\Theta(y_i)\Psi(\bH_\bma(y_i,\tau))} ~:~ 0 \le u \le
w_i^*-1, 0 \le t \le w_i-1\}$ is a set of $w_i^*w_i$ independent
traces, which can uniquely determine $\Psi(\bH_\bma(y_i,\tau))$ by
means of its dual basis. Thus, to recover $\bH_\bma(y_i,\tau)$, i.e.,
$\Psi(\bH_\bma(y_i,\tau))$ (since $\Psi(\cdot)$ is a bijection), we
only need the code symbols with evaluation points $y_j$ such that
$y_j\in\Theta\setminus\{y_i\}$ to transmit
${\tr_{\F/\F_{q_i}}}\parenv{\gamma_t v_j\Psi(\bH_\bma(y_j,\tau))}$ for
$0\leq \tau\leq r/w^*_i-1$ and $0\leq t\leq w_i-1$. Therefore, we can
recover $\bA_i$ by downloading $\frac{(w^*_i+m_1-1)r}{w^*_i}$ symbols { of $\F_q$}
from any other { $w^*_i+m_1-1$} columns.

To prove claim II, we consider each row individually, i.e., the
recovery problem for $H_{\bma, \tau}(y_i)$ for $0\leq \tau\leq
r-1$. By a similar analysis, $H_{\bma,\tau}(y_i)$ for $0\leq \tau\leq
r-1$ can be determined by the following equations:
\[
\begin{split}
{\tr_{\F_q/\F_{q_i}}}\parenv{y_i^{w^*_it}v_iy_i^ug_\Theta(y_i)H_{\bma,\tau}(y_i))}=&-\sum_{\substack{j\in[m_1+m_2]\\ j\ne i}}
{\tr_{\F_q/\F_{q_i}}}\parenv{y_i^{w^*_it}v_jy_j^ug_\Theta(y_j)H_{\bma,\tau}(y_j))}\\
=&-\sum_{\substack{j\in[m_1+m_2]\\ j\ne i}}{y_{j}^u}g_\Theta(y_j){\tr_{\F_q/\F_{q_i}}}\parenv{y_i^{w^*_it}v_jH_{\bma,\tau}(y_j))}\\
=&-\sum_{\substack{j\in[m_1+m_2]\\ j\ne i,\,y_j\in \Theta}}{y_{j}^u}g_\Theta(y_j){\tr_{\F_q/\F_{q_i}}}\parenv{y_i^{w^*_it}v_jH_{\bma,\tau}(y_j))},\\
\end{split}
\]
where $0\leq u\leq w^*_i-1$, $0\leq t\leq w_i/w_i^*-1$, and
$g_\Theta(\cdot)$ is also defined by \eqref{eqn_g_theta}. Thus, the
fact that $\{y^t_i~:~0\leq t\leq w_i-1\}$ is a basis of $\F_q$ over
$\F_i$ means that we can recover $H_{\bma,\tau}(y_i)$ for $0\leq
\tau\leq r-1$ by downloading
${\tr_{\F_q/\F_{q_i}}}(y_i^{w^*_it}v_jH_{\bma,\tau}(y_j)))$ for
$y_j\in \Theta$ and $0\leq t\leq w_i/w_i^*-1$. Namely, we can recover
$\bA_i$ by downloading $\frac{(w^*_i+m_1-1)r}{w^*_i}$ symbols {  of $\F_{q}$} from any
other $ w^*_i+m_1-1$ columns.
\end{IEEEproof}

To analyze the repair bandwidth performance for the scheme introduced
in Theorem \ref{theorem_repair_TB}, we need to modify the cut-set
bound to the case that each code symbol in the array code can be
locally repaired.

\begin{theorem}\label{theorem_cut_B_locality}
{ Let $\cC$ be an $(N,K,k=Kr;L=r+\delta-1,\delta)$-RASL code }with
$(r,L-r+1)_a$-locality in which each column
corresponds to a repair set, where $0<r\leq L-1$.  Let $D$ be an
integer with $K\leq D\leq N-1$. For any $i\in [N]$ and any $D$-subset
$\cR\subseteq [N]\setminus\{i\}$, we have
\[
B({\cC,\{i\},\cR)}\geq \frac{Dr}{D-K+1}.
\]
\end{theorem}
\begin{IEEEproof}
The fact that each column corresponds to a repair set of
$(r,L-r+1)$-locality means that the punctured code over each column
has distance at least $L-r+1$.  That is, any $r$ symbols in the $i$-th
column are capable of recovering the entire $i$-th column.  Consider
the array code $\cC'$ formed by deleting an arbitrary set of $L-r$
rows from the array code $\cC$, say the last $L-r$ rows. Since $\cC$
is an an $(N,K,k=Kr;L=r+\delta-1,\delta)$-RASL code in which each column corresponds to a repair
set, it is easy to check that $\cC'$ is also an $[N,K,N-K+1]_q$ MDS
array code, but with sub-packetization $r$.
Even though each column of $\cC'$ is only a part of the corresponding
column in $\cC$, due to the locality, each column may compute its
original column from $\cC$. Thus, any repair procedure on $\cC$ may
also be run on $\cC'$, and therefore we have $B(\cC,\{i\},\cR)\geq
B(\cC',\{i\},\cR)$ for any $i\in [N]$ and any $D$-subset
  $\cR\subseteq [N]\setminus\{i\}$.  Now the desired result follows
  from Theorem \ref{theorem_cut_B}, that is, for any $i\in [N]$ and
  any $D$-subset $\cR\subseteq [N]\setminus\{i\}$, we have
\begin{equation*}
{B(\cC,\{i\},\cR)\geq B(\cC',\{i\},\cR)}\geq \frac{Dr}{D-K+1}.
\end{equation*}
\end{IEEEproof}

\begin{remark}
In the proof of Theorem \ref{theorem_cut_B_locality}, we used
${B(\cC,\{i\},\cR)\geq B(\cC',\{i\},\cR)}$. In fact, the reverse inequality,
$B(\cC,i,\cR)\leq B(\cC',i,\cR)$ is trivially true since each column of $\cC'$ is a part of the original column from $\cC$, and therefore any repair procedure running on $\cC'$ may be run on $\cC$.
\end{remark}

Based on Theorems \ref{theorem_repair_TB} and \ref{theorem_cut_B_locality},
we have the following corollary:

\begin{corollary}\label{coro_TB_node}
Let $q_0$ be a prime power, and $\F_q=\F_{q_0}(y_1,y_2,\dots,y_{m})$,
$\F_{q_i}=\F_{q_0}(y_1,\dots,y_{i-1},y_{i+1},\dots,y_{m})$.  Define
$w_i\eqdef [\F_q:\F_{q_i}]$ for all $i\in[m]$. Furthermore, let
$m_2\geq \max\{2,r\}$, where $m=m_1+m_2$ and $k=m_1r$. If $0<r\leq
\min\{w_1,\,w_2,\,\dots,\,w_{m}\}$, and $\gcd(r,w_i)=1$ for all
$i\in[m]$, then for any $i\in [m]$ we can recover $\bA_i$ by
downloading $r+m_1-1$ symbols in $\F_q$ from any other $r+m_1-1$ { racks (columns)}, which is
exactly the optimal bandwidth with respect to the cut-set bound of
Theorem \ref{theorem_cut_B_locality}.
\end{corollary}
\begin{IEEEproof}
According to Theorem \ref{theorem_repair_TB}-I, the facts that
$\gcd(w_i,r)=1$ for $i\in [m]$ and $r\leq
\min\{w_1,\,w_2,\,\dots,\,w_{m}\}$ mean that $w^*_i=r$ and we can
recover $\bA_i$ by downloading $r+m_1-1$ symbols from any other
$r+m_1-1$ columns.  Note that $\cC$ is an $(m,m_1,k=m_1r;L=r+\delta-1,\delta)$-RASL code  in which each column corresponds a repair set,
where $0<r\leq L-1$.  By Theorem \ref{theorem_cut_B_locality},
$$B(\cC,{\{i\}},m_1+r-1)\geq \frac{(r+m_1-1)r}{r}.$$
Therefore, in this case, the code $\cC$ has optimal repair bandwidth.
\end{IEEEproof}

{ \begin{remark}
For multiple rack erasures, the method used in Theorem \ref{theorem_repair_TB} may not
achieve the optimal bandwidth by Theorem \ref{theorem_cut_B_locality}. The key aspect may be to design a repair scheme that allows some downloaded data to be repeatedly used when repairing multiple rack erasures, similar to the method used in \cite{tamo2018repair}.
Finding optimal schemes for repairing multiple rack-erasures is still an open problem.
\end{remark}}

\section{Repairing Tamo-Barg codes: Partial erasures}\label{sec-partial}
In the previous section, we demonstrated that Tamo-Barg
codes may exhibit optimal repair properties when represented in an
array form. In this section, we further explore the partial repair
problem for the rack-aware system with locality, i.e.,
array codes under the assumption that each column of the
array is an $(r,\delta)$-repair set. Specifically, we consider the
scenario where some repair sets have failed, meaning that certain
columns contain more than $\delta-1$ erasures. We seek to determine
the minimum amount of data that needs to be downloaded from $D$ remaining
columns, and how to construct a code that achieves the minimum repair
bandwidth for this model.

To begin, we provide some necessary definitions for the partial erasure case in Section \ref{sec-rack-aware-system}.

\begin{definition}
{ Let $\cC$ be an $(N,K,k=Kr;L=r+\delta-1,\delta)$-RASL code}, where $0<r\leq L-1$.  Let
$\cI=\{i_1,i_2,\ldots, i_\tau\}\subseteq [N]$ denote the failed
columns, and let $E_{i_t}\subseteq [L]$ with $|E_{i_t}|\geq \delta$
for $t\in[\tau]$ denote the corresponding erasures in the
$i_t$-th column.  For a $D$-subset $\cR\subseteq [N]\setminus \cI$,
define $B(\cC,\cI,\cE,\cR)$ as the \emph{minimum repair bandwidth} for
$\{c_{i,j} ~:~ i \in \cI, j \in E_i\}$, i.e., the smallest number of
symbols of $\F_q$ helper { racks (columns)} need to send in order to recover the
erasure pattern $\cE=\{E_{i_t} ~:~ t\in[\tau]\}$ (where each
helper { rack (column)} $j\in \cR$ may send symbols that depend solely on
$c_j\in\F_q^L$).

\end{definition}

\begin{figure}[tb]
  \centering
\scalefont{0.87}
\begin{tikzpicture}[scale=0.87]
   \draw [|<->|, blue] (-7,4) -- (-1,4);
   \node (code)  at (-4,4) [above,black]{codewords (evaluation domain)};
    \matrix [nodes={text width=12mm,draw},nodes={minimum size=8mm}](Ori_code) at (-4,0)
    {\node{$c_{1,1}$};&\node{$c_{1,2}$};&\node{$\cdots$};
    &\node{$c_{1,m}$};\\
    \node{$c_{2,1}$};&\node{$c_{2,2}$};&\node{$\cdots$};
    &\node{$c_{2,m}$};\\
    \node{$\vdots$};&\node{$\vdots$};&\node{$\cdots$};&\node{$\vdots$};\\
    \node{$c_{r,1}$};&\node{$c_{r,2}$};&\node{$\cdots$};&\node{$c_{r,m}$};\\
    \node[fill=light_gray]{$c_{r+1,1}$};&\node[fill=light_gray]{$c_{r+1,2}$};&\node[fill=light_gray]{$\cdots$};&\node[fill=light_gray]{$c_{r+1,m}$};\\
    \node[fill=light_gray]{$\vdots$};&\node[fill=light_gray]{$\vdots$};&\node[fill=light_gray]{$\cdots$};&\node[fill=light_gray]{$\vdots$};\\
    \node[fill=light_gray]{$c_{r+\delta-1,1}$};&\node[fill=light_gray]{$c_{r+\delta-1,2}$};&\node[fill=light_gray]{$\cdots$};&\node[fill=light_gray]{$c_{r+\delta-1,m}$};\\
    };

    \draw [|<->|, blue] (10,4) -- (0,4);
    \node (coef) at (5,4) [above,black]{polynomials (coefficient domain)};

    \matrix [nodes={text width=22mm,draw},nodes={minimum size=8mm}] at (5,0)
    {\node{$e^{0}_1=h_1(y_1)$};&\node{$e^0_2=h_1(y_2)$};&\node{$\cdots$};
    &\node{$e^0_m=h_1(y_m)$};\\
    \node{$e^1_1=h_2(y_1)$};&\node{$e^1_2=h_2(y_2)$};&\node{$\cdots$};
    &\node{$e^1_m=h_2(y_m)$};\\
    \node{$\vdots$};&\node{$\vdots$};&\node{$\cdots$};&\node{$\vdots$};\\
    \node{$e^{r-1}_1=h_{r}(y_1)$};&\node{$e^{r-1}_2=h_{r}(y_2)$};
    &\node{$\cdots$};&\node{$e^{r-1}_m=h_{r}(y_m)$};\\
    };
    \draw [blue] (-7,-3)--(-6,1);
    \draw [blue] (-7,1)--(-6,-3);
    \draw [blue] (0.5,-1.5)--(1.5,1.5);
    \draw [blue] (0.5,1.5)--(1.5,-1.5);

    \draw [|<->|, blue] (10.6,1.8) -- (10.6,0.9) node at (10.6,1.35) [right, blue]{$[m,k/r]$ GRS code};
    \node at (10.6,0) [right, blue]{$\vdots$};
    \draw [|<->|, blue] (10.6,-1.8) -- (10.6,-0.9) node at (10.6,-1.35) [right, blue]{$[m,k/r]$ GRS code};
    \draw [|<->|, blue] (-0.5,-2) -- (2.1,-2);
    \node (f1) at (0.8,-2) [below, blue]{$f_{\bm a,1}(x)$};

    \matrix [nodes={text width=35mm,draw},nodes={minimum size=8mm}](mat_coef) at (1,-8)
    {\node{$e^{0}_1=h_{0}(y_1)$};\\
    \node{$e^{1}_1=h_{1}(y_1)$};\\
    \node{$\vdots$};\\
    \node{$e^{|E|-\delta}_1=h_{|E|-\delta}(y_1)$};\\
    \node{$c_{j_1,1}$};\\
    \node{$\vdots$};\\
    \node{$c_{j_{r+\delta-1-|E|},1}$};\\};

    \draw [blue,->,dashed,very thick] (-6.5,2)--(mat_coef.north);
    \path[blue,->,very thick,bend right] (7,2) edge node[above] {Repairing $[m,k/r]$ GRS code} (1,2);
    \path[blue,->,very thick,bend left](code.north)edge node[above] {Lemma \ref{lemma_F_a}} (coef.north);
	\path[blue,->,very thick,dashed](f1.east)edge node[above] {} (mat_coef.north);
    \path[blue,->,bend left,very thick](mat_coef.west)edge node[above] {recovered} (-6.5,-3);

\end{tikzpicture}

  \caption{Partial repairing of Tamo-Barg codes. Herein, we initially reduce the repair problem of failed items in the first column to the repair problem for the coefficients of the corresponding polynomial remainder. Secondly, according to Lemma \ref{lemma_F_a}, these coefficients can be recovered by repairing certain codewords within a Reed-Solomon code. Finally, repairing these symbols of the codewords ensures the repair of the desired components.}
  \label{fig:partial_repairing}
\end{figure}
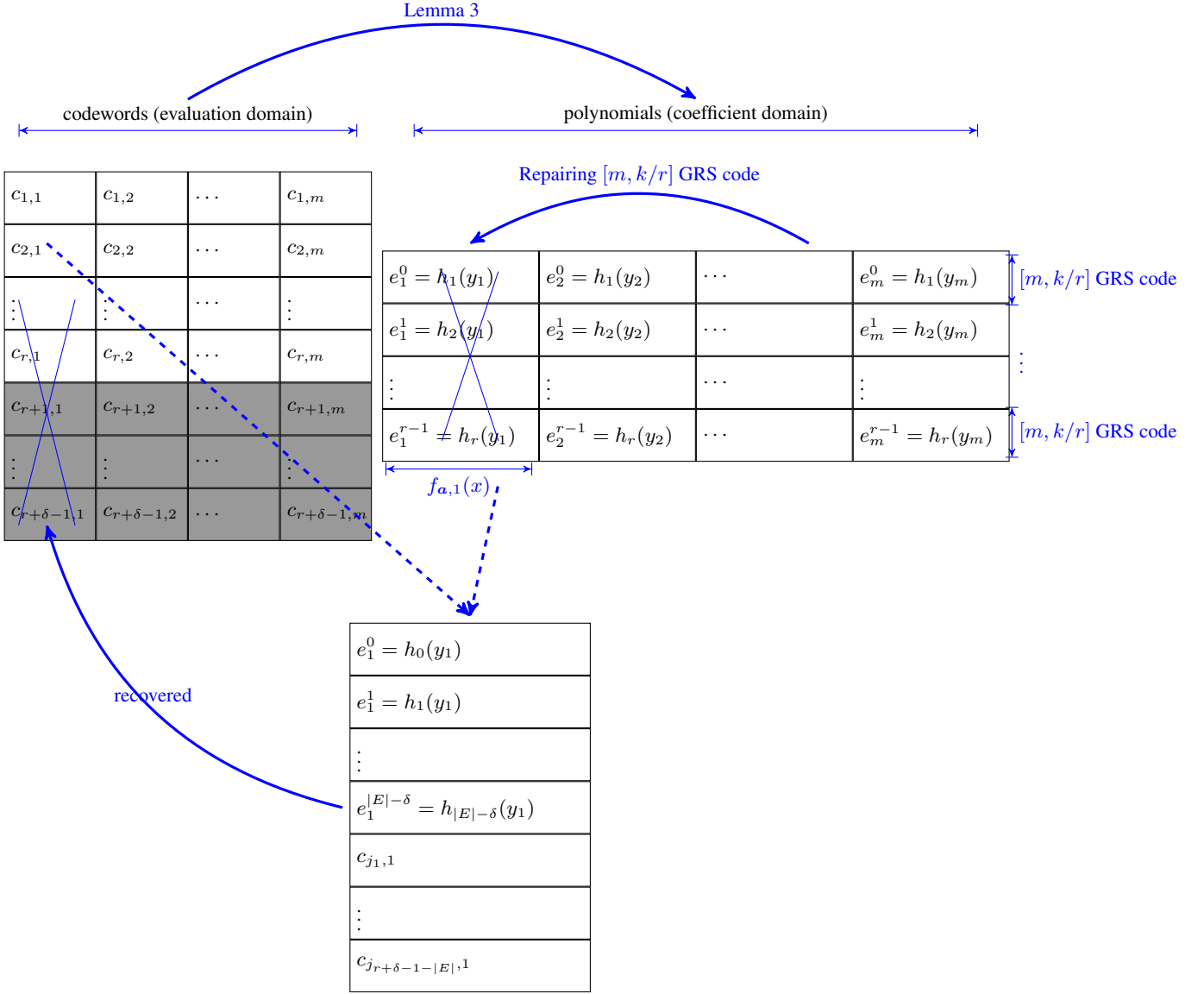

First, we consider the case where a single column contains erasures,
but unlike Theorem~\ref{theorem_repair_TB}, the column is not fully
erased. Namely, we consider the case where $\cE=\{E_{i}\}$, and $E_i=E$, i.e., { the partial
erasure case} for $\lambda=1$ in Section \ref{sec-rack-aware-system}.
The main idea of the repair process is explained in Figure \ref{fig:partial_repairing}.

\begin{theorem}\label{theorem_repair_TB_partial}
Consider the setting of Construction~\ref{cons_TB_array}. Let $q_0$ be
a prime power, and $\F_q=\F_{q_0}(y_1,y_2,\dots,y_{m})$,
$\F_{q_i}=\F_{q_0}(y_1,\dots,y_{i-1},y_{i+1},\dots,y_{m})$.  Define
$w_i\eqdef [\F_q:\F_{q_i}]$ for each $i\in[m]$. For any given set
$E\subseteq [L]$ with $\delta\leq |E|\leq L$, consider the case that
the elements in $\bA_i|_E$ are erasures.
\begin{itemize}
\item[I.] Define
  $w^*_i\triangleq\frac{L-\delta+1}{\gcd(w_i,L-\delta+1)}$.  If $w^*_i
  \le w_i$ and $m_2\geq \max\{2,w^*_i\}$, then we can recover
  $\bA_i|_E$ { by downloading $(w^*_i+m_1-1)(\abs{E}-\delta+1)$ elements of $\F_{q_i}$, i.e.,}
  $\frac{(w^*_i+m_1-1)(\abs{E}-\delta+1)}{w^*_i}$
  symbols { in $\F_q$} from any other $w^*_i+m_1-1$ { racks} (columns).
\item[II.] Assume $w^*_i$ is a positive integer. If $w^*_i|w_i$
  and $m_2\geq \max\{2,w^*_i\}$, then we can recover $\bA_i|_E$ { by
  downloading $(w^*_i+m_1-1)(\abs{E}-\delta+1)$ elements of $\F_{q_i}$, i.e.,} $\frac{(w^*_i+m_1-1)(\abs{E}-\delta+1)}{w^*_i}$ symbols { in $\F_q$} from
  any other $w^*_i+m_1-1$ { racks} (columns).
\end{itemize}
\end{theorem}

\begin{IEEEproof}
By Construction \ref{cons_TB_array}, for recovering of $\bA_i|_E$, it
is sufficient to recover $f_{\bma,i}(x)\equiv
F_\bma(x)\pmod{h(x)-y_i}$.  By Lemma \ref{lemma_F_a}, for $i\in[m]$,
\begin{equation*}
F_\bma(x)\pmod{h(x)-y_i}= f_{\bma,i}(x)=\sum_{j=0}^{r-1}H_{\bma,j}(y_i)x^{j},
\end{equation*}
where $\deg(H_{\bma,j}(x))\leq m_1-1$.  Recall that in this case we
have $\delta=L-r+1$, and we know $r+\delta-1-|E|$ elements of $\bA_i$,
i.e., those elements in $\bA_i|_{[r+\delta-1]\setminus E}$.  Thus, one
sufficient condition for repairing $f_{\bma,i}(x)$ is to recover
$H_{\bma,j}(y_i)$ for $0\leq j\leq |E|-\delta$. Note that for these
$j$'s, $H_{\bma,j}(y_t)$ for $t\in[m]\setminus\{i\}$ can be
calculated by the elements in $\bA_t$.

Define the following array $\cA_1$ from the Tamo-Barg code:
\begin{equation*}
\cA_1=\bracenv{A^*_\bmb ~:~
{A^*_\bmb}=\parenv{\bA^*_1,\bA^*_2,\dots,\bA^*_m}
=\parenv{F_\bmb(\beta_{u,v})}_{(r+\delta-1)\times m}
,\,\,{\rm for \,\,any}\,\,\bmb\in \F^{m_1(|E|-\delta+1)}_q},
\end{equation*}
where
\begin{equation*}
F_\bmb(x) \pmod{h(x)-y_i} = f^*_{\bmb,i}(x)={\sum_{t=0}^{|E|-\delta}}b_{i,t}x^t
\end{equation*}
and $\bmb=(b_{1,0},b_{1,1},\dots,b_{1,|E|-\delta},b_{2,0},\dots,b_{m_1,|E|-\delta}).$
For any $\bma\in \F^{rm_1}_q$, let $V(\bma)=(v_{1,0},v_{1,1},\dots,v_{1,|E|-\delta},v_{2,0},\dots,v_{m_1,|E|-\delta})$
with $v_{i,j}=H_{\bma,j}(y_{i})$ for $i\in[m_1]$ and $0\leq j\leq |E|-\delta$,
then for any $m_1+1\leq i\leq m$ we have
\begin{equation*}
F_{V(\bma)}(x)\equiv \sum_{0\leq j\leq |E|-\delta}v_{i,j}x^j\pmod{h(x)-y_i},
\end{equation*}
{where}
\begin{equation*}
{v_{i,j}=H_{\bma,j}(y_{i})}\quad \text{for }m_1+1\leq i\leq m,\,\,0\leq j\leq |E|-\delta,
\end{equation*}
according to Lemma \ref{lemma_F_a}.  Now consider the repair problem
of the $i$-th column $\bA^*_i$ of the codeword ${A^*_{V(\bma)}}\in
\cA_1$. Since the repair problem is equivalent to recovering the
coefficients of $f^*_{V(\bma),i}(x)$, i.e., $v_{i,j}=H_{\bma, j}(y_i)$
for $0\leq j\leq |E|-\delta$, it is also a sufficient condition to
recover $A_i|_E$.  Thus, the bandwidth of repairing $\bA_i|_E$ is
upper bounded by the bandwidth of repairing $\bA^*_i$ for
${\bA^*_{V(\bma)}}$. Now the desired results follows from Theorem
\ref{theorem_repair_TB}.
\end{IEEEproof}

We now move on to address the challenge of repairing erasure patterns
in scenarios where multiple repair sets (columns) have failed,
specifically when $|E_{i_t}|\geq \delta$ for $t\in[\tau]$, i.e.,
the partial erasure case with $\lambda=\tau>1$.

Consider a given erasure pattern
$\cE=\{E_{i_1},E_{i_2},\dots,E_{i_{\tau}}\}$ with $|E_{i_t}|\geq
\delta$ for all $t\in[\tau]$. Define
\begin{equation}
  \label{eqn_theta_i}
  \begin{split}
\Lambda_{\cE}(x)&\eqdef \prod\limits_{E_{j}\in\cE}(x-y_j)=\prod_{t\in[\tau]}(x-y_{i_t}),\\
\Omega_{t,\cE,\alpha}(x)&\eqdef\frac{\tr_{\F_q/\F_{q_1}}\parenv{\alpha\Lambda_{\cE}(x)}}{(x-y_{i_t})\prod_{E_{i_j}\in\cE,\,j\ne t}(y_{i_t}-y_{i_j})}.
  \end{split}
\end{equation}

\begin{theorem}
Let $\cC$ be the code generated by Construction \ref{cons_TB_array},
and $\cE=\{E_{i_1},E_{i_2},\dots,E_{i_{\tau}}\}$ be an erasure pattern
with $|E_{i_t}|\geq \delta$ for all $t\in[\tau]$. For a subfield
$\F_{q_1}\subset \F_{q}$, if $m_2\geq \frac{q\tau}{q_1}$ then the
erasure pattern can be recovered { by downloading $M$ elements of $\F_{q_1}$, i.e.,
 with repair bandwidth $\frac{M}{\ell}$}
by contacting all the remaining { racks (columns)}, where $M=\sum_{t\in[\tau]}|E_{i_t}|-\tau(\delta-1)$ and
$q=q_1^{\ell}$.

\end{theorem}
\begin{IEEEproof}
Without loss of generality, assume $i_t=t$, namely, we consider the case
$\cE=\{E_1,E_2,\dots,E_{\tau}\}$, $|E_i|\geq \delta$ for all
$i\in[\tau]$. Let $w=\max\{|E_i| ~:~ i\in[\tau]\}$ and for any $1\leq
\ell\leq w-\delta+1$ define
\[
R_\ell\eqdef \{i\in[\tau] ~:~ |E_i|\geq \delta-1+\ell\}.
\]
By Construction \ref{cons_TB_array}, for recovering of $\bA_i|_{E_i}$,
$E_i\in \cE$, one possible method is to recover $f_{\bma,i}(x)=
F_\bma(x)\pmod{h(x)-y_i}$.  By Lemma \ref{lemma_F_a}, for $i\in[m]$,
\begin{equation*}
F_\bma(x)\pmod{h(x)-y_i}= f_{\bma,i}(x)=\sum_{j=0}^{r-1}H_{\bma,j}(y_i)x^{j},
\end{equation*}where $\deg(H_{\bma,j}(x))\leq m_1-1$.
Note that $(H_{\bma,j}(y_1),H_{\bma,j}(y_2),\cdots,H_{\bma,j}(y_m))$
can be regarded as a codeword of an $[m=m_1+m_2,m_1,m_2+1]_q$ GRS
code, i.e.,
$(H_{\bma,j}(y_1),H_{\bma,j}(y_2),\cdots,H_{\bma,j}(y_m))\in
\GRS_{m_1}(\bm 1,\bm Y)$, where $\bm Y=(y_1,y_2,\dots,y_m)$.

First, we are going to apply the trace function and
$\Omega_{i,\cE,\alpha}(x)$ defined by \eqref{eqn_theta_i} to recover
$H_{\bma,j}(y_i)$ for all $j\in[w-\delta+1]$ and $i\in R_\ell$. For
$i\in [m]\setminus R_1$, $\bA_{i}$ suffers from at most $\delta-1$
erasures, i.e., at least $r$ components are accessible. Hence, { rack (column)}
$i$, $i\in [m]\setminus R_1$, can recover
$f_{\bma,i}(x)=\sum_{j=0}^{r-1}H_{\bma,j}(y_i)x^{j}$ since
$\deg(f_{\bma,i}(x))<r$.  For $j\in[w-\delta+1]$ and $i\in R_j$,
the { rack (column)} $i$ downloads
\[
\tr_{\F_{q}/\F_{q_1}}\parenv{
\frac{-H_{\bma,j}(y_t)}{y_t-y_i}}
\]
from the { racks (columns)} $t$, for $t\in [m]\setminus R_1$. Since the dual of a
GRS code is a GRS code with the same code locators, there exists a
vector $\bm \Theta=(\theta_1,\theta_2,\dots,\theta_m)\in \F^{m}_q$
such that $\GRS_{m_1}^{\bot}(\bm 1,\bm Y)=\GRS_{m_2}(\bm \Theta,\bm
Y)$. Since $m_2\geq \frac{q\tau}{q_1}$ and
$\deg(\Omega_{i,\cE,\alpha}(x))\leq \frac{q\tau}{q_1}-1$ we have
that
\[(\theta_1\Omega_{i,\cE,\alpha}(y_1),\theta_2\Omega_{i,\cE,\alpha}(y_2),\dots,\theta_{m}\Omega_{i,\cE,\alpha}(y_m))\in
\GRS_{m_2}(\bm \Theta, \bm Y)=\GRS_{m_1}^{\bot}(\bm 1,\bm Y).
\]
Thus, for $\ell\in[w-\delta+1]$ and $i\in R_\ell$, we have
\begin{equation*}
0=\sum_{t\in[m]}\theta_t\Omega_{i,\cE,\alpha}(y_t) H_{\bma,r-j}(y_t) \text{ for $\alpha \in \F_q$.}
\end{equation*}
By \eqref{eqn_theta_i}, we have $\Omega_{i,\cE,\alpha}(y_i)=\alpha$
for $t\in R_1\setminus\{i\}$ and $\Omega_{i,\cE,\alpha}(y_t)=0$ for
$i\in R_1$, which means that the preceding equation can be rewritten
as
\begin{align*}
0&=\sum_{t\in[m]}\theta_t\Omega_{i,\cE,\alpha}(y_t) H_{\bma,j}(y_t)\\
&=\theta_i\Omega_{i,\cE,\alpha}(y_i)H_{\bma,j}(y_i) +\sum_{t\in [m]\setminus R_1}\theta_t\Omega_{i,\cE,\alpha}(y_t) H_{\bma,j}(y_t)\\
&=\theta_i\alpha H_{\bma,j}(y_i) +\sum_{t\in [m]\setminus R_1}\theta_t\Omega_{i,\cE,\alpha}(y_t) H_{\bma,j}(y_t),
\end{align*}
i.e.,
\begin{equation}\label{eqn_recover}
\begin{split}
\theta_i\alpha H_{\bma,j}(y_i)= -\sum_{t\in [m]\setminus R_1}\theta_t\Omega_{i,\cE,\alpha}(y_t) H_{\bma,j}(y_t)
\end{split}
\end{equation}
for $\ell\in[w-\delta+1]$, $i\in R_\ell$, and $\alpha\in \F_q$. Let
$\{\alpha_1,\alpha_2,\dots, \alpha_T\}$ be a basis of $\F_q$ over
$\F_{q_1}$ and $\{\beta_1,\beta_2,\dots,\beta_T\}$ be its dual
basis. By \eqref{eqn_recover}, for $j\in[w-\delta+1]$, $i\in R_j$, and
$v\in[T]$,
\begin{align*}
\tr_{\F_q/\F_{q_1}}\parenv{\theta_i\alpha_v H_{\bma,j}(y_i)}&= \tr_{\F_q/\F_{q_1}}\parenv{-\sum_{t\in [m]\setminus R_1}\theta_t\Omega_{i,\cE,\alpha_t}(y_t) H_{\bma,j}(y_t)}\\
&=\sum_{t\in [m]\setminus R_1}\tr_{\F_q/\F_{q_1}}\parenv{\frac{-\tr_{\F_q/\F_{q_1}}\parenv{\alpha_v\Lambda_{\cE}(y_t)}H_{\bma,j}(y_t)}{y_t-y_i}}\\
&=\sum_{t\in [m]\setminus R_1}\tr_{\F_q/\F_{q_1}}\parenv{\alpha_v\Lambda_{\cE}(y_t)}\tr_{\F_q/\F_{q_1}}\parenv{\frac{-H_{\bma,j}(y_t)}{y_t-y_i}},
\end{align*}
where the last two equalities hold by \eqref{eqn_theta_i} and the fact
$\tr_{q/q_1}\parenv{\alpha_v\Lambda_{\cE}(x)}\in \F_{q_1}$ for $x\in
\F_q$.  Therefore, for $\ell\in[w-\delta+1]$, $i\in R_\ell$,
$H_{\bma,j}(y_i)$ can be recovered as
\[\theta_iH_{\bma,j}(y_i)=\sum_{v\in[T]}\tr_{\F_q/\F_{q_1}}\parenv{\theta_i\alpha_v H_{\bma,j}(y_i)}\beta_v,\]
by downloading $\tr_{\F_{q}/\F{q_1}}\parenv{
  \frac{-H_{\bma,j}(y_t)}{y_t-y_i}}$ from the { racks (columns)} $t$ for $t\in
[m]\setminus R_1$. Namely, we can recover $H_{\bma,j}(y_i)$ for
$\ell\in[w-\delta+1]$ and $i\in R_\ell$, i.e., $H_{\bma,j}(y_i)$ with
$i\in[\tau]$ and $r+\delta-1-|E_i|\leq j\leq r-1$.

Secondly, for $i\in[\tau]$, we may rewrite $f_{\bma,i}(x)$ as
\[
f_{\bma,i}(x)=\sum_{j=0}^{r-1}H_{\bma,j}(y_i)x^{j}=\sum_{j=r+\delta-1-|E_i|}^{r-1}H_{\bma,j}(y_i)x^{j}+g_{\bma,i}(x),
\]
where $\deg(g_{\bma,i}(x))\leq r+\delta-|E_i|-2$. Note that the above
analysis recovers $H_{\bma,j}(y_i)$ for $i\in[\tau]$ and
$r+\delta-|E_i|-1\leq j\leq r-1$. Recall that we know
$r+\delta-1-|E_i|$ elements of $\bA_i$, i.e., those elements in
$\bA_i|_{[r+\delta-1]\setminus E_i}$. This is to say that
$\bA_i|_{[r+\delta-1]\setminus E_i}$ can recover $g_{\bma, i}(x)$ with
$\deg(g_{\bma, i}(x))\leq r+\delta-|E_i|-2$ for $i\in[\tau]$. Thus,
for $i\in[\tau]$, $f_{\bma,i}(x)\equiv F_{\bma}(x)\pmod{h(x)-y_i}$ can
be recovered. Therefore, we can recover $\bA_i$ for $i\in[\tau]$
according to Construction \ref{cons_TB_array}, which completes the
proof.
\end{IEEEproof}

When the finite field is sufficiently large, it is possible to further
decrease the partial repair bandwidth. The main idea of repairing
these erasure patterns is applying Lemma \ref{lemma_F_a} to reduce
this problem into the repair problem of a class of Reed-Solomon codes.

\begin{lemma}\label{lemma_multi_repair}
Let $\bY=(y_1,y_2,\dots,y_{m})\in\F_{q_1}^m$ and $\cC_1=\GRS_{m_1}(\bm
1,\bY)$ be an $[m=m_1+m_2,m_1,m_2+1]_{{q_1}}$ GRS code. If $\cC_1$ has
$(\tau,D)$ optimal repair property with $1\leq \tau\leq m_2$ and
$m_1<D\leq m-\tau$, then for any
$\cE=\{E_{i_1},E_{i_2},\dots,E_{i_\tau}\}$ with $E_{i_j}\subseteq [L]$
and $|E_{i_j}|=w\geq \delta$ for $j\in[\tau]$, the code $\cC$
generated by Construction \ref{cons_TB_array} satisfies
\[
B(\cC,\cI,\cE,\cR)\leq \frac{\tau D\ell(w-\delta+1)}{D-k+\tau}
\]
for any $|\cI|=\tau$ and $|\cR|=D$ with $\cR\subseteq [m]\setminus \cI$,
where $L=r+\delta-1$ and $q_1=q^{\ell}$.
\end{lemma}

\begin{IEEEproof}
According to Construction \ref{cons_TB_array}, in order to recover
$\bA_{i_t}|_{E_{i_t}}$, $i\in[\tau]$, it is sufficient to recover
$f_{\bma,i_t}(x)= F_\bma(x)\pmod{h(x)-y_{i_t}}$. Note that by
Lemma \ref{lemma_F_a}, we have
\[f_{\bma,i_t}(x)=\sum_{j=0}^{r-1} H_{\bm a,j}(y_{i_t})x^j,\]
where $\deg (H_{\bm a,j}(x))<m_1$.  Since we know $L-w$ evaluations
for each $f_{\bma, i_t}$, i.e., the values of
$f_{\bma,i_t}(\beta_{i_t,j})$ for $j\in [\ell]\setminus E_{i_t}$, we
only need to figure out $H_{\bm a,j}(y_{i_t})$ for $0\leq j\leq
w-\delta$ in order to recover $f_{\bma,i_t}(\beta_{i_t,j})$ for $j\in
E_{i_t}$. Recall that for each $i\in \cR\subseteq [m]\setminus
\cI$ we know $r+\delta-1$ evaluations for $f_{\bma, i}$, which means
we can figure out $H_{\bm a,j}(y_{i})$ for $0\leq j\leq r-1$.

Now, for any $0\leq j\leq w-\delta$, we regard $(H_{\bm
  a,j}(y_1),H_{\bm a,j}(y_2),\dots,H_{\bm a,j}(y_{m_1+m_2}))$ as a
codeword of $\cC_1=\GRS_{m_1}(\bm 1,\bY)$.  In this way, the repair
problem is reduced to $w-\delta+1$ parallel repair problems for the
code $\cC_1=\GRS_{m_1}(\bm 1,\bY)$ when the code symbols with index
$i\in \cR$ are known (helper { racks (columns)}) and the code symbols with index
$i\in \cI$ are erasures.  Since $\cC_1$ has the $(\tau,D)$ optimal
repair property, then for any $\cI\subseteq [m]$ with
$|\cI|=\tau$ and $\cR\subseteq [m] \setminus \cI$ with
$|\cR|=D$, we can recover $H_{\bm a, j}(y_i)$ for $i\in \cI$ by
{ downloading $\frac{\tau D}{D-k+\tau}$ elements over $\F_{q_1}$, i.e.,
$\frac{\tau D\ell}{D-k+\tau}$ elements over $\F_q$}. Thus,
the desired result follows since we have $w-\delta+1$ layers to
recover $H_{\bm a, j}(y_{i})$ for these $j$ and $i\in \cI$.
\end{IEEEproof}

By the preceding proof of Lemma \ref{lemma_multi_repair}, the repair
method achieves bandwidth $\frac{\tau D\ell (w-\delta+1)}{D-k+\tau}$
for the code $\cC$ constructed by Construction \ref{cons_TB_array}
only for the specified $\cI$ and $\cR$.  We remark that there are
known explicit repair methods for GRS codes such as the one
introduced in \cite{tamo2018repair} for multiple erasures.

\subsection{A lower bound on the partial-repair bandwidth}

A natural question arises regarding the partial-repair problem: what
is the theoretical bound for the partial-repair bandwidth?
{ Since the main tool to solve this problem is a kind of direct graphs,
in this section, we use vertex to denote a rack (column) of the
RASL codes. }
Let $L$
denote the number of elements stored in each node. Assume
that there is a node, say the $i$-th vertex, suffering from $L-s_i$
erasures in the positions given in $E_i$. Define $\beta(L,s_i,D)$ as
the number of elements the system needs to download from each of the
$D$ helper vertices to recover the $L-s_i$ erased elements.

Inspired by the idea of the information-flow graph presented in
\cite{dimakis2010network}, we propose a solution to this problem by
defining a special kind of information-flow graph called a
\emph{partial information-flow graph}. The basic idea is to allow each
vertex that experiences erasures to have a certain amount of surviving
information. When a vertex experiences partial erasure, the system needs
to recover the erased portion of information for the goal vertex. Since
the recovered vertex inherits the surviving information from the
original vertex, it is named as an inheritor.

\begin{definition}
A directed acyclic graph is said to be a \emph{partial
information-flow graph} if it satisfies the following:
\begin{itemize}
  \item A source vertex $S$, corresponding to the original data which
    will be stored into $N$ initial storage vertices.
  \item Initial storage vertices $X^{(i)}$, each of them consists of an
    input vertex $X^{(i)}_{in}$ and an output vertex
    $X^{(i)}_{out}$. $X^{(i)}_{in}$ and $X^{(i)}_{out}$ are connected
    by a directed edge $(X^{(i)}_{in},X^{(i)}_{out})$ with capacity
    equal to the number of elements stored at $X^{(i)}$, i.e., $r \le
    L$, where $r$ is the number of original elements stored at
    $X^{(i)}_{in}$.  $S$ connects to each $X^{(i)}_{in}$ by a directed
    edge $(S, X^{(i)}_{in})$ {with capacity $r$}.
  \item To model the dynamic behavior of storage systems such as
    erasures and repair, the time factor is also considered. At any
    given time, vertices are either active or inactive. At the initial
    time step, the storage vertices $X^{(i)}$ are all active and the
    source vertex $S$ is inactive. Later on, at any given time step, if
    a vertex suffers from a partial erasure, say the vertex
    $v=(v_{in},v_{out})$, then the vertex $v$ is set to be inactive and
    we create a direct inheritor $(I_{in},I_{out})$, which is
    connected with $v_{out}$ by an edge $(v_{out},I_{in})$ with
    capacity $s(v)$, where $s(v)$ is the number of surviving
    information symbols of the vertex $v$. The vertex $I$ also needs to
    download $\beta(L,s(v),D)$ symbols from each of $D$ other active
    vertices, i.e., we add $D$ directed edges
    $(v^{(j)}_{out},I^{(i)}_{in})$ with a capacity of
    $\beta=\beta(L,s(v),D)$. Finally, we set the vertex $I$ as active.
 \item A data collector vertex $DC$, corresponding to a request to
   construct the data.  $DC$ connects to $K$ active vertices with subscript ``out" by directed
   edges with infinite capacity to recover the original data.
\end{itemize}
\end{definition}

\begin{remark}
The partial information-flow graph is a direct generalization of the
information-flow graph to include the case of repairing partial
erasures of a vertex. Specifically, each failed vertex $v_i$ (progenitor)
will have a portion of information (denoted by $s_i$ surviving
elements) inherited by its reconstructed inheritor vertex. We also
extend the information-flow graph to include the possibility that each
vertex may contain redundancies, which corresponds to the concept of
locality in Definition~\ref{def_r_delta_i}, where each vertex is
regarded as a repair set. Therefore, when $s_i=0$ for all $i\in[N]$
and $L=r$, the partial information-flow graph reduces to the
information-flow graph defined in~\cite{dimakis2010network}.
\end{remark}

For partial information-flow graphs, two examples are presented in Fig.
\ref{fig:initiation} and \ref{fig:partialflew} to explain the initiation
and dynamic behavior of storage systems, respectively. We now use partial information-flow graphs to analyze the repair
problem of distributed storage systems in the following setting:
\begin{itemize}
\item Basic recovery assumption: The file contains $M$ symbols, and is
  stored in $N$ vertices, each of which has $L$ symbols, and any $K$
  vertices are capable of recovering all the $M$ symbols of the file.
\item $(r,L-r+1)$-locality: For each vertex, any $r\leq L$ symbols can
  recover all the symbols stored in this vertex.
\item Erasure model: There are $L-s_i$ symbols in the $i$-th
  vertex that are erased.
\end{itemize}

\begin{figure}[tb]
  \centering
\scalefont{0.5}
\begin{tikzpicture}[scale =1]
	\node[mycircle, dotted] (S) {$\bS$} ;
	\node[mycircle, above right=80pt and 50pt of S] (nodein0) {$ X^{(1)}_{in}$} ;
    \node[mycircle, above right=80pt and 100pt of S] (nodeout0) {$ X^{(1)}_{out}$} ;
	\node[mycircle, above right=40pt and 50pt of S] (nodein1) {$X^{(2)}_{in}$} ;
    \node[mycircle, above right=40pt and 100pt of S] (nodeout1) {$X^{(2)}_{out}$} ;
	\node[mycircle, above right=0pt and 50pt of S] (nodein2) {$X^{(3)}_{in}$} ;
    \node[mycircle, above right=0pt and 100pt of S] (nodeout2) {$X^{(3)}_{out}$} ;
	\node[rectangle,above right=-40pt and 75pt of S](text1){$\vdots$};
	\node[mycircle,above right=-80pt and 50pt of S] (nodein4) {$X^{(N)}_{in}$} ;
    \node[mycircle,above right=-80pt and 100pt of S] (nodeout4) {$X^{(N)}_{out}$} ;
    \node[mycircle, below right=0pt and 200pt of S] (DC) {$DC$};

	\path[black,->,very thick,auto] (S.east) edge node {$r$} (nodein0.west)
	      (S.east) edge node {$r$} (nodein1.west)
	      (S.east) edge node {$r$} (nodein2.west)
	      (S.east) edge node {$r$} (nodein4.west)
          (nodein0.east) edge node {$r$} (nodeout0.west)
	      (nodein1.east) edge node {$r$} (nodeout1.west)
          (nodein2.east) edge node {$r$} (nodeout2.west)
          (nodein4.east) edge node {$r$} (nodeout4.west)
          (nodeout4.east) edge node {$+\infty$} (DC.west)
          (nodeout1.east) edge node {$+\infty$} (DC.west)
          (nodeout2.east) edge node {$+\infty$} (DC.west);
	
\end{tikzpicture}

  \caption{An example of the the initial status of the information flow graph,
  where one possible data collector $DC$ is included.}
  \label{fig:initiation}
\end{figure}
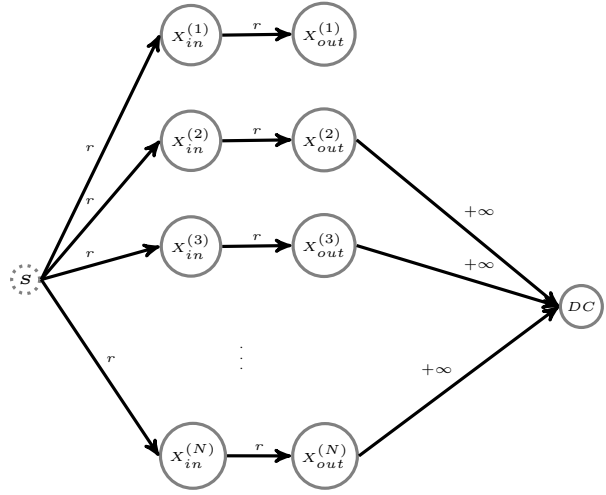
\begin{figure}
  \centering
\scalefont{0.5}
\begin{tikzpicture}[scale =1]
	\node[mycircle,dotted] (S) {$\bS$} ;
	\node[mycircle, dotted, above right=80pt and 50pt of S] (nodein0) { $ X^{(1)}_{in}$} ;
    \node[circle split, dotted, draw, above right=80pt and 100pt of S] (nodeout0) { $s_1$ \nodepart{lower}  $ X^{(1)}_{out}$} ;
	\node[mycircle, dotted, above right=40pt and 50pt of S] (nodein1) { $X^{(2)}_{in}$} ;
    \node[mycircle, dotted, above right=40pt and 100pt of S] (nodeout1) { $X^{(2)}_{out}$} ;
    \node[rectangle,above right=20pt and 75pt of S](D1){$\vdots$};
	\node[mycircle, above right=0pt and 50pt of S] (nodein2) { $X^{(i)}_{in}$} ;
    \node[mycircle, above right=0pt and 100pt of S] (nodeout2) { $X^{(i)}_{out}$} ;
	\node[rectangle,above right=-40pt and 75pt of S](D2){$\vdots$};
	\node[mycircle,above right=-80pt and 50pt of S] (nodein4) { $X^{(N)}_{in}$} ;
    \node[mycircle,above right=-80pt and 100pt of S] (nodeout4) { $X^{(N)}_{in}$} ;
	\node[circle split, dotted, draw, right=40pt of nodeout0] (hin0) { $s_1$ \nodepart{lower}  $ I^{(1)}_{in}$} ;
    \node[mycircle, dotted, right=20pt of hin0] (hout0) { $ I^{(1)}_{out}$} ;
    \node[circle split, draw, right=40pt of hout0] (hin1) { $s_1$ \nodepart{lower}  $ I^{(2)}_{in}$} ;
    \node[mycircle, right=20pt of hin1] (hout1) { $ I^{(2)}_{out}$} ;
    	\node[circle split, draw, right=40pt of nodeout1] (hin2) { $s_2$ \nodepart{lower}  $ I^{(3)}_{in}$} ;
    \node[mycircle, right=20pt of hin2] (hout2) { $ I^{(3)}_{out}$} ;

    \node[mycircle, below right=10pt and 40pt of hout1] (DC) {$DC$};

	\path[black,->,very thick,auto] (S.east) edge node {$r$} (nodein0.west)
	      (S.east) edge node {$r$} (nodein1.west)
	      (S.east) edge node {$r$} (nodein2.west)
	      (S.east) edge node {$r$} (nodein4.west)
          (nodein0.east) edge node {$r$} (nodeout0.west)
	      (nodein1.east) edge node {$r$} (nodeout1.west)
          (nodein2.east) edge node {$r$} (nodeout2.west)
          (nodein4.east) edge node {$r$} (nodeout4.west)
          (hin0.east) edge node {$r$} (hout0.west)
          (hin1.east) edge node {$r$} (hout1.west)
          (hin2.east) edge node {$r$} (hout2.west)
          (nodeout0.north) edge [bend left,dotted,red] node {$s_1$} (hin0.north)
          (hout0.north) edge [bend left,dotted,red] node {$s_1$} (hin1.north)
          (nodeout1.north) edge [bend left,dotted,red] node {$s_2$} (hin2.north)
          (hout1.east) edge [bend left] node {$+\infty$} (DC.west)
          (hout2.east) edge [bend right] node {$+\infty$} (DC.west)
          (nodeout2.east) edge [bend right] node {$+\infty$} (DC.west);

          \path[black,->,dotted,very thick,auto]
          (5.5,4) edge node {$\beta$} (hin0.west)
          (5.5,3.5) edge node {$\beta$} (hin0.west)
          (5.5,3) edge node {$\vdots$} (hin0.west)
          (5.5,2.5) edge node {$\beta$} (hin0.west)
          (9.5,4) edge node {$\beta$} (hin1.west)
          (9.5,3.5) edge node {$\beta$} (hin1.west)
          (9.5,3) edge node {$\vdots$} (hin1.west)
          (9.5,2.5) edge node {$\beta$} (hin1.west)
          (5.5,2.2) edge node {$\beta$} (hin2.west)
          (5.5,1.7) edge node {$\beta$} (hin2.west)
          (5.5,1.5) edge node {$\vdots$} (hin2.west)
          (5.5,1) edge node {$\beta$} (hin2.west);

	\draw [|-|, very thick,blue] (13.8,2) arc  (240:120:15pt);
    \draw node[blue] at (13.8,1.5) {${\bm Kr}$};

\end{tikzpicture}

  \caption{An example for three inheritors in the partial
  information flow graph with parent $X^{(i)}$, $i\in\{1,2\}$, where the first node suffers partial erasures twice and the second node is erased once. Here we use dotted vertices to denote
  inactive nodes. We also include a possible data collector $DC$ as an example.}
  \label{fig:partialflew}
\end{figure}
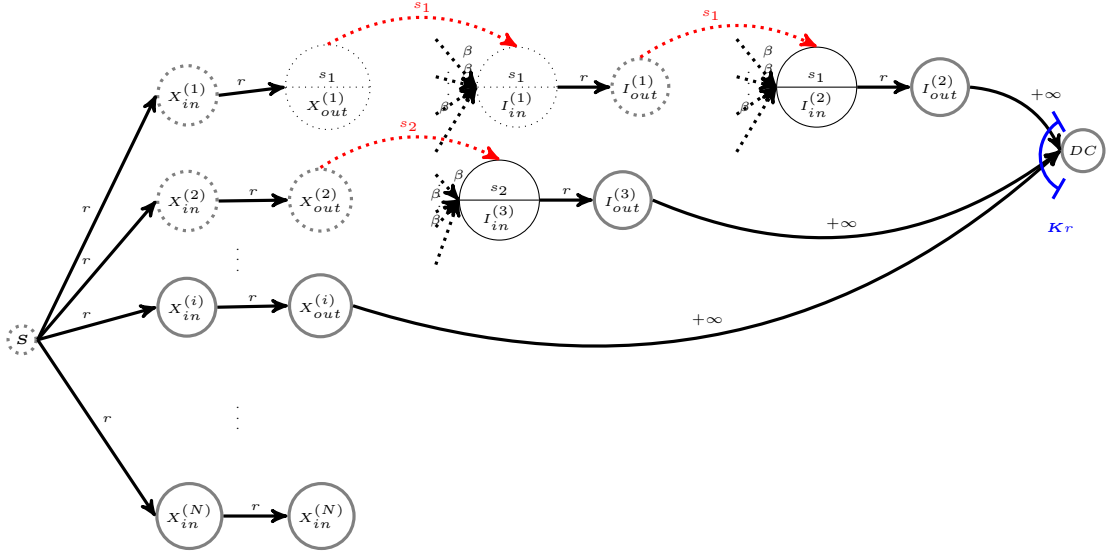

Given positive integers $N>K$, $D\leq N-1$, $L\geq r$, $s_i\leq L$ for
all $i\in [N]$ and a real number $\beta\geq 0$, let
$G(N,K,r,D,\beta;\bms=(s_1,s_2,\dots,s_N))$ denote a family of partial
information-flow graphs with all possible evolutions. The parameter
tuple $(N,K,r,D,\beta;\bms)$ is said to be \emph{feasible} if there
exists a locally repairable array code with repair bandwidth $\beta$
and sub-packetization $L$, with $L-s_i$ erasures in the $i$-th vertex.

\begin{proposition}\label{prop_inheritor}
Let $G\in G(N,K,r,D,\beta;\bms=(s_1,s_2,\dots,s_N))$ be the partial
information-flow graph for a given time step. Then the following hold:
\begin{itemize}
  \item For any $v\in G$, there exists a unique list of vertices
    $B(v)=\{v^{(1)}=v,v^{(2)},\cdots,v^{(t)}\}$ { such that} $v^{(i+1)}$ is an
    inheritor of $v^{(i)}$ for $i\in [t-1]$, where $t$ denotes
    the amount of inheritors in $G$ for vertex $v$.
  \item For any pair of vertices $v, v'\in G$, if $v'\not\in B(v)$ and
    $v\not\in B(v')$ then there is at most one edge
    {$(v^{(i)}_{out},v_{in}')$ for $v^{(i)}_{out}\in \widetilde{B}(v)$}
    and $v_{in}'\in \widetilde{B}(v')$, where
    $\widetilde{B}(v)=\{v_{in},v_{out} ~:~ v=(v_{in},v_{out})\in B(v)\}$.
\end{itemize}
\end{proposition}

\begin{IEEEproof}
For each vertex $v$ in $G$, if it has a direct inheritor, then that
inheritor is unique. We can then define $B(v)$ as the set of all
inheritors (direct or indirect) of $v$ in $G$, which is also
unique. The size of $B(v)$, denoted by $|B(v)|$, represents the number
of inheritors of $v$ in $G$, including $v$ itself.

For the second part, we assume that there exists an edge
$({v^{(i)}_{out}},v_{in}')$ for ${v^{(i)}}\in B(v)$.  Since $v'\not\in
B(v)$ and $v\not\in B(v')$, we can conclude that $v^{(i)}$ is active
when $v'$ is included in $G$. This also means that $v^{(j)}$ for
$j\in[i-1]$ are inactive, i.e., $(v^{(j)}_{out},v'_{in})\not\in
E$. Note that $v^{(i)}$ being active implies that $v^{(j)}$ for $j>i$
have not been included yet, i.e., $(v^{(j)}_{out},v'_{in})\not\in
E$. This completes the proof.
\end{IEEEproof}

\begin{definition}
A \emph{cut} in the partial information flow graph $G$ between $S$ and
$DC$ is a subset of edges $W$ such that each directed path from $S$ to
$DC$ contains at least one edge in $W$. Furthermore, the minimal cut
is the cut with {the smallest} edge capacity sum.  We define the
capacity of $W$ as
\[
C(W)=\sum_{e\in W} C(e),
\]
where $C(e)$ denotes the capacity of an edge $e$.
\end{definition}

\begin{theorem}\label{theorem_bandwidth}
For given positive integers $N>K$, $D\leq N-1$, $L\geq r$, $s_i\leq L$
for $i\in [N]$, the parameter tuple $(N,K,D,L,\beta;\bms)$ is feasible
if and only if $Kr\leq {c(N,K,D,\beta;\bms)}$ under a large enough
finite field, where $c(N,K,D,\beta;\bms)$ satisfies
\[
c(N,K,D,\beta;\bms)  =
\sum_{i=0}^{\min\{K,D\}-1}\min\{(D-i)\beta+s_{j_i},r\}+\sum_{i=\min\{K,D\}}^{K-1} \min\{s_{j_{i}},r\},
\]
where $\beta\geq 0$ is a real number, and $s_{j_0}\leq s_{j_1}\leq
\dots\leq s_{j_{N-1}}$.
\end{theorem}

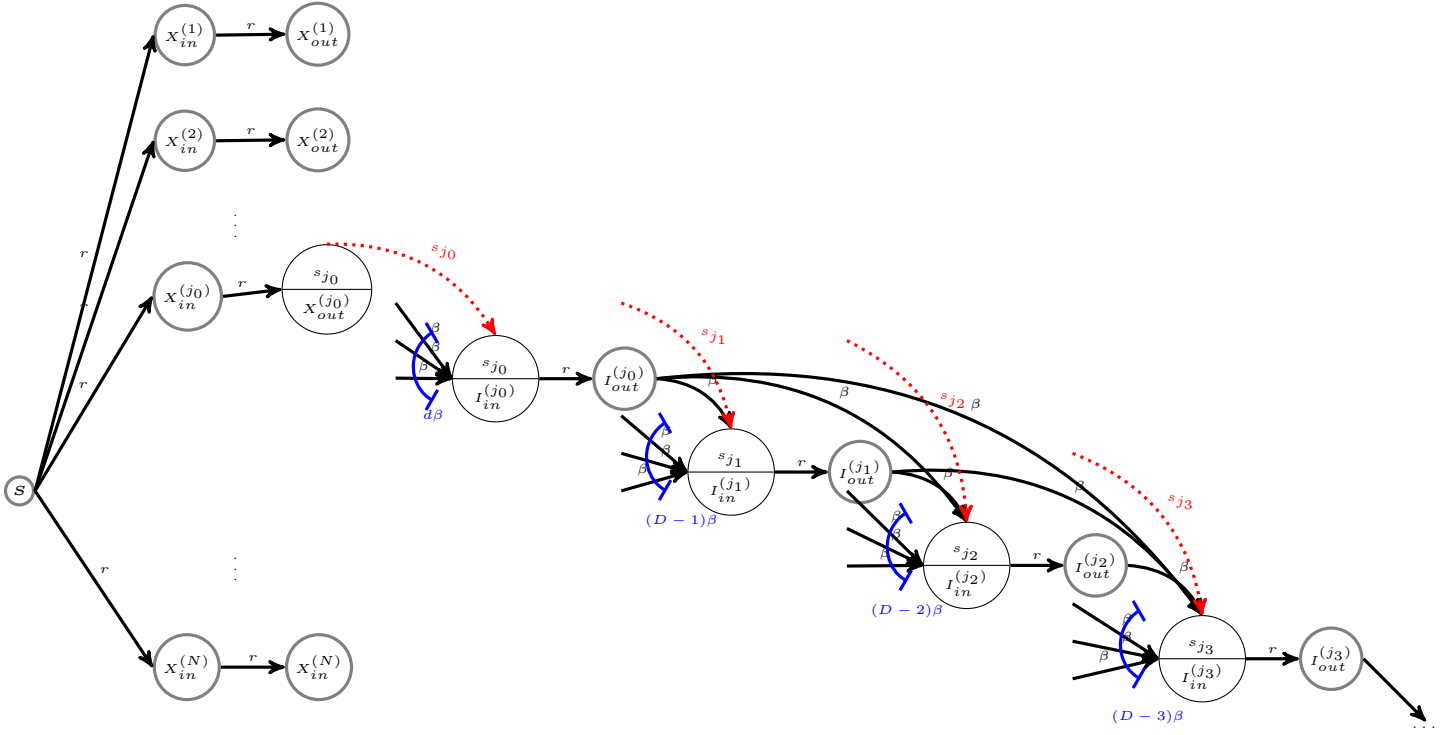
\begin{figure*}[tb]
  \centering
\tikzsetnextfilename{2-butterflies}
\scalefont{0.5}
\begin{tikzpicture}[scale =1]
	\node[mycircle] (S) {$\bS$} ;
    \node[mycircle, above right=160pt and 50pt of S] (nodein1+) { $ X^{(1)}_{in}$} ;
    \node[mycircle, above right=160pt and 100pt of S] (nodeout1+) { $X^{(1)}_{out}$} ;
	\node[mycircle, above right=120pt and 50pt of S] (nodein2+) { $X^{(2)}_{in}$} ;
    \node[mycircle, above right=120pt and 100pt of S] (nodeout2+) { $X^{(2)}_{out}$} ;
    \node[rectangle,above right=90pt and 75pt of S](text2){$\vdots$};
	\node[mycircle, above right=60pt and 50pt of S] (nodein0) { $ X^{(j_0)}_{in}$} ;
    \node[circle split, draw, above right=60pt and 100pt of S] (nodeout0) { $s_{j_0}$ \nodepart{lower}  $ X^{(j_0)}_{out}$} ;
	\node[rectangle,above right=-40pt and 75pt of S](text1){$\vdots$};
	\node[mycircle,above right=-80pt and 50pt of S] (nodein4) { $X^{(N)}_{in}$} ;
    \node[mycircle,above right=-80pt and 100pt of S] (nodeout4) { $X^{(N)}_{in}$} ;
	\node[circle split, draw, below right=10pt and 40pt of nodeout0] (hin0) { $s_{j_0}$ \nodepart{lower}  $ I^{(j_0)}_{in}$} ;
    \node[mycircle, right=20pt of hin0] (hout0) { $ I^{(j_0)}_{out}$} ;
    \node[circle split, draw, below right=15pt and 20pt of hout0] (hin1) { $s_{j_1}$ \nodepart{lower}  $ I^{(j_1)}_{in}$};
    \node[mycircle, right=20pt of hin1] (hout1) { $ I^{(j_1)}_{out}$} ;
    \node[circle split, draw, below right=15pt and 20pt of hout1] (hin2) { $s_{j_2}$ \nodepart{lower}  $ I^{(j_2)}_{in}$};
    \node[mycircle, right=20pt of hin2] (hout2) { $ I^{(j_2)}_{out}$} ;
    \node[circle split, draw, below right=15pt and 20pt of hout2] (hin3) { $s_{j_3}$ \nodepart{lower}  $ I^{(j_3)}_{in}$};
    \node[mycircle, right=20pt of hin3] (hout3) { $ I^{(j_3)}_{out}$} ;
    \node[below right=15pt and 20pt of hout3, black] (Dot) {$\dots$} ;

	\path[black,->,very thick,auto] (S.east) edge node {$r$} (nodein0.west)
	      (S.east) edge node {$r$} (nodein1+.west)
           (S.east) edge node {$r$} (nodein2+.west)
	      (S.east) edge node {$r$} (nodein4.west)
          (nodein0.east) edge node {$r$} (nodeout0.west)
          (nodein1+.east) edge node {$r$} (nodeout1+.west)
          (nodein2+.east) edge node {$r$} (nodeout2+.west)
          (nodein4.east) edge node {$r$} (nodeout4.west)
          (hin0.east) edge node {$r$} (hout0.west)
          (hin1.east) edge node {$r$} (hout1.west)
          (hin2.east) edge node {$r$} (hout2.west)
          (hin3.east) edge node {$r$} (hout3.west)
          (hout0.east) edge [bend left] node {$\beta$} (hin1.north)
          (hout0.east) edge [bend left] node {$\beta$} (hin2.north)
          (hout0.east) edge [bend left] node {$\beta$} (hin3.north)
          (hout1.east) edge [bend left] node {$\beta$} (hin2.north)
          (hout1.east) edge [bend left] node {$\beta$} (hin3.north)
          (hout2.east) edge [bend left] node {$\beta$} (hin3.north)
          (nodeout0.north) edge [bend left,dotted,red] node {$s_{j_0}$} (hin0.north)
          (5,2.5) edge node {$\beta$} (hin0.west)
          (5,2) edge node {$\beta$} (hin0.west)
          (5,1.5) edge node {$\beta$} (hin0.west)
          (8,2.5) edge [bend left,dotted,red] node {$s_{j_1}$} (hin1.north)
          (8,1) edge node {$\beta$} (hin1.west)
          (8,0.5) edge node {$\beta$} (hin1.west)
          (8,0) edge node {$\beta$} (hin1.west)
          (11,2) edge [bend left,dotted,red] node {$s_{j_2}$} (hin2.north)
          (11,0) edge node {$\beta$} (hin2.west)
          (11,-0.5) edge node {$\beta$} (hin2.west)
          (11,-1) edge node {$\beta$} (hin2.west)
          (14,0.5) edge [bend left,dotted,red] node {$s_{j_3}$} (hin3.north)
          (14,-1.5) edge node {$\beta$} (hin3.west)
          (14,-2) edge node {$\beta$} (hin3.west)
          (14,-2.5) edge node {$\beta$} (hin3.west);

	\draw [|-|, very thick,blue] (5.5,1.2) arc  (240:120:15pt);
    \draw node[blue] at (5.5,1) {$\small d\beta$};
    \draw [|-|, very thick,blue] (8.6,0) arc  (240:120:15pt);
    \draw node[blue] at (8.8,-0.4) {$\small (D-1)\beta$};
    \draw [|-|, very thick,blue] (11.8,-1.2) arc  (240:120:15pt);
    \draw node[blue] at (11.8,-1.6) {$\small (D-2)\beta$};
    \draw [|-|, very thick,blue] (14.9,-2.5) arc  (240:120:15pt);
    \draw node[blue] at (15,-3) {$\small (D-3)\beta$};
	\draw [->,very thick](hout3.east) -- (Dot.north);

\end{tikzpicture}

  \caption{Each inheritor receives $s_j$ surviving symbols from its
    parent, which does not cost network traffic, and is denoted with a
    dotted arrow. We also highlight the inheritance by dividing the
    node into two parts, where the one labeled with $s_j$ corresponds
    to the inheritance.}
  \label{fig:minicut}
\end{figure*}

Before going to the proof of Theorem \ref{theorem_bandwidth}, we
recall the well known theorem about minimum cut and maximum flow for
network coding, and translate it into our partial information-flow
graph setting.

\begin{lemma}\label{lemma_cut_flew}
For a partial information-flow graph $G$, a data collector $DC$ can
recover the original file if and only if the minimum cut between the source
$S$ and a data collector $DC$ is larger than or equal to the file size
under a large enough finite field.
\end{lemma}
\begin{IEEEproof}
Regard $DC$ as a terminal, then the partial information-flow graph $G$
can be viewed as a broadcast network with source $S$ and original file
with size $M$. Then, according to the well known cut-set
bound~\cite{cover1999elements} for network coding, under a large
enough finite field the network has a linear
solution~\cite{ho2006random}, i.e., $DC$ can get enough information to
recover the original file if and only if the minimum cut between the
source $S$ and the terminal $DC$ is larger than or equal to the file
size $M$.
\end{IEEEproof}

\begin{IEEEproof}[Proof of Theorem~\ref{theorem_bandwidth}]
According to Lemma \ref{lemma_cut_flew}, proving this theorem is
equivalent to proving that the minimum cut between the source vertex $S$
and the data collector $DC$ is larger than or equal to the file size
$M=Kr$ if and only if $Kr\leq c(N,K,D,\beta;\bms),$ for any $G\in
G(N,K,r,D,\beta;\bms)$.

In one direction, we show that there exists a partial information-flow
graph $G_1\in G(N,K,r,D,\beta;\bms)$ such that the minimum cut
$W(G_1)$ between the source $S$ and the data collector $DC$ is at most
$c(N,K,D,\beta;\bms)$, i.e., $C(W(G_1))\leq {c(N,K,D,\beta;\bms)}$.
Consider a sequence of {inheritor vertices} denoted as
$(I^{(j_t)}_{in},I^{(j_t)}_{out})$ for $0\leq t\leq K-1$ such that
there exists a directed edge
$(I^{(j_{t_1})}_{out},I^{(j_{t_2})}_{in})$ with capacity $\beta$ if
and only if $0< t_2-t_1\leq D$. Now consider the data collector that
connects to the vertices $(I^{(j_t)}_{in},I^{(j_t)}_{out})$ for $0\leq
t\leq K-1$. It is easy to check that the minimum cut satisfies
\begin{equation*}
C(W(G_1))\leq \begin{cases}
\sum_{i=0}^{K-1}\min\{(D-i)\beta+s_{j_i},r\}, &\text{if $K\leq D$,}\\
\sum_{i=0}^{D-1}\min\{(D-i)\beta+s_{j_i},r\}+\sum_{i=D}^{K-1}\min\{s_{j_i},r\},&\text{if $K>D$,}
\end{cases}
\end{equation*}
as illustrated in Fig. \ref{fig:minicut}. Thus, $G_1$ is feasible only if
\[\sum_{i=0}^{\min\{K,D\}-1}\min\{(D-i)\beta+s_{j_i},r\}+\sum_{i=\min\{D,K\}}^{K-1}\min\{s_{j_{i}},r\}\geq Kr.\]

In the other direction, we claim that for any $G_2\in
G(N,K,r,D,\beta;\bms)$ with {$D$ helper vertices, any} cut $W$ between
{the source $S$ and a data collector} $DC$ should satisfy
\[
C(W)\geq \sum_{i=0}^{\min\{K,D\}-1}\min\{(D-i)\beta+s_{j_i},r\}+\sum_{i=\min\{D,K\}}^{K-1}\min\{s_{j_{i}},r\}.
\]
To prove this we assume the $K$ active vertices connected with $DC$ are
$\{Y^{(i)}_{out} ~:~ i\in [K]\}$ and the cut $W$ partitions the vertices
into two subsets, $U$ and $\overline{U}$, such that $S\in U$ and $DC
\in \overline{U}$. Let $\overline{U}_{out}$ denote the set of {vertices
  with subscript ``out"}. We only need to consider the case $\{Y^{(i)}_{out} ~:~ i\in [K]
\}\subseteq \overline{U}_{out}$ since the capacity between $Y^{(i)}_{out}$ and $DC$ is infinite which implies
$C(W)=+\infty$ for the case $Y^{(i)}_{out}\in U$.  Since the graph $G_2$ is a directed
acyclic graph then it has a topological sorting which means the vertices
can be ordered such that the edge $(v_i,v_j)$ implies that $i<j$. We
assume the set $\{Y^{(i)}_{out} ~:~ i\in [K] \}\cap \overline{U}$ has
size $\tau$ and is topologically ordered as
$(Y^{(i_1)}_{out},Y^{(i_2)}_{out},\cdots,Y^{(i_\tau)}_{out})$, where
$\tau\leq K$. Now, we find a list of vertices
$(L^1_{out},L^2_{out},\cdots,L^K_{out})$ satisfying the following two
conditions:
\begin{itemize}
  \item [I] $L_{out}^1$ is the first vertex of $\overline{U}_{out}$
    under the topological sorting.
  \item [II] For $i\in[K]$, $L_{out}^i$ is the first vertex of
    $\overline{U}_{out}\setminus
    (\bigcup_{j=1}^{i-1}\widetilde{B}(L^{j}))$ under the topological
    sorting.
\end{itemize}

For $L_{out}^1$, if $L^1_{in}\not\in \overline{U}$ then
$(L^1_{in},L_{out}^1)\in W$ with capacity $r$.  If $L^1_{in}\in
\overline{U}$ then, by Condition I, $W$ contains all the input edges
of $L^{1}_{in}$.  Those edges have capacity $r$ in total if $L^1$ does
not have a progenitor, or capacity $s_{t_1}+D\beta$ otherwise.

For $1<i\leq K$ and $L^i_{in}\not\in \overline{U}$, we have
$(L^i_{in},L_{out}^i)\in W$ with capacity $r$.

For $1<i\leq K$ and $L^i_{in}\in \overline{U}$, we have the following
two subcases:
\begin{itemize}
  \item $L^i$ has no progenitor, i.e., $L^i$ is one of the original
    storage vertices connected with the source $S$. In that case, we have
    $(S,L^i_{in})\in W$ with capacity $r$.
  \item $L^i$ has a progenitor, say $L$. In that case, we claim that
    $L_{out}\in U$.  Otherwise, by Proposition
    \ref{prop_inheritor}-(1), the fact $L^i\in
    \overline{U}_{out}\setminus
    (\bigcup_{j=1}^{i-1}\widetilde{B}(L^{j}))$ means $L_{out}\in
    \overline{U}_{out}\setminus
    (\bigcup_{j=1}^{i-1}\widetilde{B}(L^{j}))$, which contradicts the
    fact that $L^i$ is the first vertex of $\overline{U}_{out}\setminus
    (\bigcup_{j=1}^{i-1}\widetilde{B}(L^{j}))$ under the the
    topological sorting. Similarly, $W$ should contain all the input
    edges of $L^i_{in}$ except for those edges from vertices belonging to
    $\widetilde{B}(L^{j})$ for $j\in[i-1]$. By Proposition
    \ref{prop_inheritor}-(2), there are at most $i-1$ edges coming
    from the vertices of $\widetilde{B}(L^{j})$ for $j\in[i-1]$ and each
    of them should have capacity $\beta$. Thus, $W$ should contain
    input edges of $L^i_{in}$ with total capacity at least $s_{t_i}$
    if $i>D$ and $s_{t_i}+(D-i+1)\beta$ otherwise. Herein, we
    highlight that $t_i\not\in \{t_1,t_2,\dots,t_{i-1}\}$ since
    $L^i\in \overline{U}_{out}\setminus
    (\bigcup_{j=1}^{i-1}\widetilde{B}(L^{j}))$.
\end{itemize}

Therefore, for any cut $W$ between $S$ and $DC$, the capacity satisfies
\[C(W)\geq {\sum_{i=0}^{\min\{K,D\}-1}}\min\{(D-i)\beta+s_{j_i},r\}+{\sum_{i=\min\{D,K\}}^{K-1}}\min\{s_{j_{i}},r\},\]
where the last inequality holds by the fact $s_{j_1}\leq s_{j_2}\leq
\dots \leq s_{j_{N}}$, i.e.,
\[\sum_{i=1}^{K}s_{t_i}\geq \sum_{i=1}^{K}s_{j_i}\]
for any possible $t_i$ with $0\leq i\leq K-1$.
Thus, we have
\[C_{\min}=C(W(G_1))=\sum_{i=0}^{\min\{K,D\}-1}\min\{(D-i)\beta+s_{j_i},r\}+{\sum_{i=\min\{D,K\}}^{K-1}}\min\{s_{j_{i}},r\},\]
where $C_{\min}$ denotes the minimum cut capacity of all the graphs
$G\in G(N,K,r,D,\beta;\bms)$.
{ For an example of the exact minimum cut, the reader may refer to Fig. \ref{fig:minicut}.}
 Therefore, $C_{\min}\geq Kr$ is
equivalent with
$\sum_{i=0}^{\min\{K,D\}-1}\min\{(D-i)\beta+s_{j_i},r\}+{\sum_{i=\min\{D,K\}}^{K-1}}\min\{s_{j_{i}},r\}\geq
Kr$.
\end{IEEEproof}

As the next step, we study the relationship between $r$ and
$N,K,\beta,D,M$ by solving the inequality $Kr\leq c(N,K,D,\beta;\bms)$
in Theorem~\ref{theorem_bandwidth}.

\begin{theorem}\label{theorem_cut_set}
Let $\bms=(s_1,s_2,\cdots,s_{N})$ and $s_{j_0}\leq s_{j_1}\leq
\dots\leq s_{j_{N-1}}<s_{j_{N}}\triangleq +\infty$.  For $1\leq t\leq
K-1$, define
\begin{align*}
s^{*}_{j_t}&\eqdef\frac{\sum_{i=0}^{t-1}s_{j_i}+(K-t)s_{j_t}}{K},\\
g(t)&\eqdef\sum_{i=0}^{t-1}s_{j_i}+{t(D-K)}\beta+\frac{t(t+1)\beta}{2},
\end{align*}
and for $0\leq t\leq K-1$,
\begin{equation}\label{eqn_f}
f(t)\eqdef\frac{1}{K(D+1-K)+\frac{t(t+1)}{2}+t(K-t-1)}.
\end{equation}
When $s_{j_{\tau-1}}<r\leq s_{j_{\tau}}$ for $\tau\in[N]$, the
parameter tuple $(N,K,r,D,\beta;\bms)$ is feasible if and only if
$r\geq r^*(N,K,D,\beta,\bms)$ and the solution can be achieved via
linear codes, where
\[
r^*(N,K,D,\beta;\bms)=
\begin{cases}
\frac{M}{K}, & \beta\in [f(0)K(r-s_{j_0}),\infty),\\
\frac{M-g(t)}{K-t}, &\beta\in (f(t)K(r-s^*_{j_t}),f(t-1)K(r-s^*_{j_{t-1}})],\\
\end{cases}
\]
for $t\in[\tau^*-1]$ with $\tau^*\triangleq \min\{\tau,K\}$.
If $r\leq s_{j_0}$ then the {parameter tuple $(N,K,r,D,\beta;\bms)$} is always feasible for any $\beta\geq 0$.
\end{theorem}

\begin{IEEEproof}
By Theorem \ref{theorem_bandwidth}, we need to determine the exact threshold by considering fixed values of $\beta$, $\bms$, and $D\geq K$, and minimizing {$r=r_{\min}(\beta,\bms,D)$} such that
\[
\sum_{i=0}^{K-1}\min\{(D-i)\beta+s_{j_i},r\}\geq M.
\]
Let
\begin{equation}\label{eqn_b_i}
b_i=(D-(K-1-i))\beta+s_{j_i} \quad\text{for }0\leq i\leq K-1
\end{equation}
and $b_{K}=+\infty$.
Let $\tau$ be the integer such that $s_{j_{\tau-1}}< r\leq s_{j_\tau}$ with $1\leq \tau\leq N$, we have
\[
C(r)\triangleq\sum_{i=0}^{K-1}\min\{(D-i)\beta+s_{j_i},r\}=\begin{cases}
Kr, & 0\leq r<b_0,\\
(K-t)r+\sum_{0\leq i\leq t-1}b_i, & b_{t-1}\leq r<b_{t},\,\, 1\leq t\leq \tau^*=\min\{\tau,K\},\\
\end{cases}
\]
where $b_0\leq b_1\leq \dots \leq b_{K-1}$ by \eqref{eqn_b_i}.  The
{preceding} equality means that $C(r)$ is strictly increasing from $0$
to $(K-\tau^*)r+\sum_{i=0}^{\tau^*-1}b_i$ as $r$ increasing from $0$
to $b_{\tau^*-1}$. Thus, to find a minimum value for $r$ such that
$C(r)\geq M$, we only need to calculate $C^{-1}(M)$ when $M\leq
(K-\tau^*)r+\sum_{i=0}^{\tau^*-1}b_i$. That is, for $1\leq t\leq
\tau^*-1$
\begin{equation}\label{eqn_CM-1}
C^{-1}(M)=\begin{cases}
\frac{M}{K}, & M\in [0,Kb_0),\\
\frac{M-g(t)}{K-t}, & M\in [\sum_{i=0}^{t-2}b_i+(K-t+1)b_{t-1},\sum_{i=0}^{t-1}b_i+(K-t)b_t),\\
\end{cases}
\end{equation}
where
\begin{equation*}
\sum_{i=0}^{t-1}b_i=\sum_{i=0}^{t-1}s_{j_i}+{t(D-K)}\beta+\frac{t(t+1)\beta}{2}=g(t).
\end{equation*}
Recall that for $1\leq t\leq K$,
\begin{equation}\label{eqn_ft}
\begin{split}
&\sum_{i=0}^{t-1}b_i+(K-t){b_{t}}\\
=&\sum_{i=0}^{t-1}s_{j_i}+{t(D-K)}\beta+{\frac{t(t+1){\beta}}{2}}+(K-t)(s_{j_t}+(D+1-K+t)\beta)\\
=&\sum_{i=0}^{t-1}s_{j_i}+(K-t)s_{j_t}+K(D+1-K)\beta+\frac{t(t+1)\beta}{2}+t(K-t-1)\beta\\
=& Ks^*_{j_t}+\frac{\beta}{f(t)},
\end{split}
\end{equation}
where $s^{*}_{j_t}\triangleq\frac{\sum_{i=0}^{t-1}s_{j_i}+(K-t)s_{j_t}}{K}$ {and} the last equality holds by \eqref{eqn_f}.
Note that $M=Kr$. Combining \eqref{eqn_ft} and \eqref{eqn_CM-1},
\[\sum_{i=0}^{t-2}b_i+(K-t+1)b_{t-1}\leq M<\sum_{i=0}^{t-1}b_i+(K-t)b_t\]
is equivalent with
\[K(r-s^*_{j_t})f(t)< \beta\leq K(r-s^*_{j_{t-1}})f(t-1)\]
for $1\leq t\leq K$, which results in the claim for the case
$r>s_{j_0}$. Finally, note that for the case $r\leq s_{j_0}\leq
\dots\leq s_{j_{N-1}}$, we have $C(r)\equiv Kr=M$, which means for
this case the parameter tuple $(N,K,r,D,\beta; \bms)$ is always
feasible.
\end{IEEEproof}

\begin{remark}
For the case $L>s_{j_0} \geq r$, i.e., the case that the number of
erased symbols is less than $L-r$, we know that $\beta=0$ is
sufficient to repair those erasures since we have $(r,L-s+1)$-locality
inside each vertex.
\end{remark}

\begin{remark}
For the case $s_{j_0}=s_{j_1}=\cdots=s_{j_{N-1}}=0$ and $L=r$, i.e.,
the ordinary case without locality, the bounds in Theorems
\ref{theorem_bandwidth} and \ref{theorem_cut_set} are exactly the
cut-set bound described in \cite{dimakis2010network} and
\cite{hu2016double} for the rack-aware model.
\end{remark}

Considering the regular case, $s_{1}=s_2=\cdots=s_{N}$, we have the
following corollary that is implied directly from
Theorem~\ref{theorem_cut_set}.

\begin{corollary}\label{coro_partial_cut_bound}
{ Let $\cC$ be an $(N,K,k=Kr;L=r+\delta-1,\delta)$-RASL code}, where
each column is a repair set with $(r,\delta=L-r+1)$-locality.  Let $D$
be an integer satisfying $K<D\leq N-1$. For any $i\in [N]$, $E_i\subseteq [L]$,
and $D$-subset $\cR\subseteq [N]\setminus\{i\}$, we have
\[
B(\cC,\{i\},\{E_i\},\cR)\geq
\begin{cases}
\frac{D(|E_i|-\delta+1)}{D-K+1}, &\text{if }|E_i|\geq \delta,\\
0, &\text{otherwise}.
\end{cases}
\]
\end{corollary}
\begin{IEEEproof}
By setting $s_1=s_2=\dots=s_N=L-|E_i|,$ if $L-|E_i|<r$, i.e., $|E_i|\geq \delta$,
then we have
\[
\beta \geq \frac{K(|E_i|-\delta+1)}{K(D-K+1)}=\frac{|E_i|-\delta+1}{D-K+1}
\]
by Theorem \ref{theorem_cut_set}, since $M=Kr$.  Thus, we have
$B(\cC,\{i\},\{E_i\},\cR)=D\beta\geq \frac{D(|E_i|-\delta+1)}{D-K+1}$
when $|E_i|\geq \delta$.  If $|E_i|<\delta$, then by Theorem
\ref{theorem_cut_set} we have $\beta\geq 0$, i.e.,
$B(\cC,\{i\},\{E_i\},\cR)\geq 0$
\end{IEEEproof}

Thus, similarly to Corollary \ref{coro_TB_node}, we have the following
conclusion for partial repairing for Tamo-Barg codes, which follows
directly from Corollary \ref{coro_partial_cut_bound} and Theorem
\ref{theorem_repair_TB_partial}.

\begin{corollary}
Consider the setting of Construction~\ref{cons_TB_array}. Let $q_0$ be
a prime power, and $\F_q=\F_{q_0}(y_1,y_2,\dots,y_{m})$,
$\F_{q_i}=\F_{q_0}(y_1,\dots,y_{i-1},y_{i+1},\dots,y_{m})$.  Define
$w_i\eqdef [\F_q:\F_{q_i}]$ for each $i\in[m]$. For any given erasure
set $E\subseteq [L]$ with $\delta\leq |E|\leq L$, we can recover
$\bA_i|_{E}$ by downloading $\frac{(r+m_1-1)(|E|-\delta+1)}{r}$
symbols { in $\F_q$} from any other $r+m_1-1$ { racks (columns)}, which is exactly the optimal
bandwidth with respect to the cut-set bound according to Corollary
\ref{coro_partial_cut_bound}.
\end{corollary}

\section{Conclusion}\label{sec-conclusion}

In this paper, we explored the rack-aware systems with locality.
Under this model, by arranging each repair set as a rack, we considered repairing
erasures that extend beyond the locality for locally repairable
codes. We presented two repair
schemes to minimize the repair bandwidth for Tamo-Barg
codes under the rack-aware model by designating each repair set as a
rack. One of the schemes achieved optimal repair for a single rack
erasure. Additionally, the cut-set bound was established for locally
repairable codes under the rack-aware model, and our repair schemes
were proven to be optimal with respect to this bound.

Moreover, we also studied the partial-repair problem for locally
repairable codes under the rack-aware model with locality,
and we introduced repair
schemes and bounds for this scenario. However, research on the
partial-repair problem for locally repairable codes is still in its
early stages.
{ In general, the partial-repair problem for locally repairable codes may be a three-dimensional one where each rack (column) contains
$L$ symbols with locality, and each symbol is a vector over a given base field, say $\F_q$.
This general problem is still widely open.}
 Furthermore, it remains an open question how to repair known
locally repairable codes such as maximally recoverable codes (partial
MDS codes), sector-disk codes, and locally repairable codes with
super-linear length. We leave these questions for future work.

\section*{Acknowledgments}

The authors are very grateful to the anonymous reviewers,
and the Associate Editor, Prof. Itzhak Tamo, for their valuable comments
that improved the quality and presentation of this paper.

\bibliographystyle{IEEEtranS}

\bibliography{HanBib}

\end{document}